\documentclass[11pt]{article}

\pdfoutput=1

\usepackage[T1]{fontenc}
\usepackage[latin9]{inputenc}
\usepackage[a4paper]{geometry}
\usepackage[active]{srcltx}
\usepackage{amsmath}
\usepackage{amssymb}
\usepackage{esint}
\usepackage{ulem}

\usepackage{xcolor}

\makeatletter
\usepackage{textcomp}

\pdfoutput=1 % if your are submitting a pdflatex (i.e. if you have
\usepackage{jheppub}% for details on the use of the package, please
\newcommand*\xbar[1]{%
  \hbox{%
    \vbox{%
      \hrule height 0.5pt % The actual bar
      \kern0.3ex%         % Distance between bar and symbol
      \hbox{%
        \kern-0.0em%      % Shortening on the left side
        \ensuremath{#1}%
        \kern-0.0em%      % Shortening on the right side
      }%
    }%
  }%
}

\usepackage{amsfonts}

\usepackage{bbm}

\newcommand{\be}{\begin{equation}}
\newcommand{\ee}{\end{equation}}
\newcommand{\bea}{\begin{eqnarray}}
\newcommand{\eea}{\end{eqnarray}}

\title{\boldmath A note on the asymptotic symmetries of electromagnetism}

\author[a]{Oscar Fuentealba,}
\author[a,b]{Marc Henneaux,}
\author[c]{and C\'{e}dric Troessaert}

\affiliation[a]{Universit\'e Libre de Bruxelles and International Solvay Institutes, ULB-Campus Plaine CP231, B-1050 Brussels, Belgium}
\affiliation[b]{Coll\`ege de France, 11 place Marcelin Berthelot, 75005 Paris, France}
\affiliation[c]{Haute-Ecole Robert Schuman, Rue Fontaine aux M\^{u}res, 13b, B-6800, Belgium}

\emailAdd{oscar.fuentealba@ulb.be}
\emailAdd{marc.henneaux@ulb.be}
\emailAdd{cedric.troessaert@hers.be}

\preprint{}

\abstract{We extend the asymptotic symmetries of electromagnetism in order to consistently include angle-dependent $u(1)$ gauge transformations $\epsilon$ that involve terms growing at spatial infinity linearly and logarithmically in $r$, $\epsilon \sim a(\theta, \varphi) r + b(\theta, \varphi) \ln r + c(\theta, \varphi)$. The charges of the logarithmic $u(1)$ transformations are found to be conjugate to those of the $\mathcal O(1)$  transformations (abelian algebra with invertible central term) while those of the $\mathcal O(r)$  transformations are conjugate to those of the subleading $\mathcal O(r^{-1})$  transformations.  Because of this structure, one can decouple the angle-dependent $u(1)$ asymptotic symmetry from the Poincar\'e algebra, just as in the case of gravity: the generators of these internal transformations are Lorentz scalars in the redefined algebra.  This implies in particular that one can give a definition of the angular momentum which is free from $u(1)$ gauge ambiguities. The change of generators that brings the asymptotic symmetry algebra to a direct sum form involves non linear redefinitions of the charges.  Our analysis is Hamiltonian throughout and carried at spatial infinity.}

\makeatother

\begin{document}
\maketitle \flushbottom

\section{Introduction}

In a recent paper \cite{Fuentealba:2022xsz}, we extended the Hamiltonian formulation of Einstein theory in the asymptotically flat context by allowing more flexible boundary conditions that involved logarithmic terms.  This generalization led to an enlargement of the asymptotic symmetry, from the original BMS$_4$ algebra \cite{Bondi:1962px,Sachs:1962wk,Sachs:1962zza} to the log-BMS$_4$ algebra, which contains, besides the familiar angle-dependent supertranslations, angle-dependent logarithmic supertranslations \cite{Bergmann:1961zz,AshtekarLog85,BeigSchmidt82,Beig:1983sw,Ashtekar:1990gc,Compere:2011ve,Troessaert:2017jcm}.  This extension was carried out at spatial infinity following the approach developed in  \cite{Henneaux:2018hdj,Henneaux:2019yax}.

The presence in the asymptotic symmetry algebra of logarithmic supertranslations had a dramatic impact, in that it enabled one to completely disentangle the Poincar\'e subalgebra from the supertranslations (ordinary and logarithmic), realizing at spatial infinity a mechanism similar to the one described at null infinity in \cite{Mirbabayi:2016axw,Bousso:2017dny,Javadinezhad:2018urv,Javadinezhad:2022hhl} (see also \cite{Chen:2021szm,Chen:2021zmu,Compere:2021inq} in that context).

The purpose of this note is to carry out the analogous construction in the electromagnetic case, where angle-dependent $u(1)$ gauge transformations play the role of the supertranslations \cite{He:2014cra,Kapec:2015ena,Campiglia:2015qka,Campiglia:2017mua,Strominger:2017zoo}.  We consistently enlarge the boundary conditions of \cite{Henneaux:2018gfi} in such a way that the asymptotic symmetries contain angle-dependent gauge transformations that grow at spatial infinity not only logarithmically in $r$, but also linearly in $r$, $$\epsilon \sim a(\theta, \varphi) r + b(\theta, \varphi) \ln r + c(\theta, \varphi) + o(1).$$
The term linear in $r$ is included because it is associated with the subleading soft theorems \cite{Lysov:2014csa,Campiglia:2016hvg,Conde:2016csj}.  

In the standard approach where $\epsilon$ is restricted to take the form $\epsilon \sim  c(\theta, \varphi) + o(1)$, the angle-dependent $u(1)$ asymptotic transformations of order one transform in a non-trivial representation of the Poincar\'e algebra \cite{Barnich:2013sxa}, \cite{Henneaux:2018gfi}. As in the case of gravity, we show that the enlargement of the symmetry enables one to disentangle the internal asymptotic angle-dependent $u(1)$ symmetries from the Poincar\'e algebra: the asymptotic symmetry algebra is the direct sum of the two.  In retrospect, this result is 
 perhaps not too surprising as it is in the line of the Coleman-Mandula theorem \cite{Coleman:1967ad}  (even though the hypotheses of this theorem are not all fulfilled).  The fact that the internal $u(1)$ improper gauge symmetries commute with the Lorentz transformations leads to an angular momentum that is free from ambiguities under asymptotic angle-dependent $u(1)$ transformations.
 
 The change of generators that brings the asymptotic symmetry algebra to a direct sum form involves  redefinitions of the Poincar\'e generators by the addition of field-dependent gauge transformations corresponding to a specific non linear redefinitions of the charges.

Our paper is organized as follows.  In Section {\bf \ref{sec:AsymAction}}, we give the form of the new, more flexible, boundary conditions and verify the finiteness of the action. We then describe the asymptotic gauge symmetries in  Section {\bf \ref{sec:Improper}}. Poincar\'e invariance is established in Section {\bf \ref{sec:PoincInv}}  and further discussed in Appendix {\bf \ref{AppendixA}}.  The algebra of the asymptotic symmetries and the redefinitions that bring it to a direct sum form are successively analysed in Sections {\bf \ref{sec:ASA}} and {\bf \ref{sec:ASAu(1)}}.  In the concluding Section {\bf \ref{sec:Conclusions}}, we outline some potential future developments.

\section{Action and asymptotic conditions}
\label{sec:AsymAction}

\subsection{Extended Hamiltonian action}

In the formulation where the conjugate momentum to $A_0$ is kept, the phase space of the Maxwell theory is spanned by the components $(A_i, A_0)$ of the vector potential and their conjugate momenta $ (\pi^i, \pi^0)$.   These are subject to the ``primary constraint''
\be \pi^0 \approx 0 \label{ref:Const-2a}\,,\ee
and Gauss's law, which arises as a ``secondary constraint'',
\be
\mathcal{G} = -\partial_{i}\pi^{i} \approx 0\,. \label{ref:Const-2b}
\ee
It is customary to eliminate $\pi^0$ to obtain a reduced theory where the canonical variables are $(A_i, \pi^i)$, the temporal component $A_0$ appearing then as a Lagrange multiplier for the (``secondary'') constraint $\mathcal{G}  \approx 0$.   We shall refrain from doing so here because $A_0$ does carry degrees of {freedom} at infinity when one imposes boundary conditions that are invariant under an angle-dependent $u(1)$ symmetry \cite{Henneaux:2018hdj,Henneaux:2018gfi}.  It is in that case useful to keep its conjugate momentum in the Hamiltonian description.

The extended Hamiltonian action of Maxwell theory is given by
\begin{align}
I_{H}[A_{i},\pi^{i},A_{0},\pi^{0};\psi,\lambda]=& \int dt  \int d^{\text{3}}x\big[\pi^{i}\dot{A}_{i}+ \pi^0\dot{A}_0- (\mathcal{H} + \tilde \epsilon \mathcal G + \tilde \mu \pi^0)  - \psi \mathcal{G}- \lambda \pi^0 \big]  +\mathfrak{B} \, , \label{eq:Action-Extended0} \\
 \mathcal H =& \frac{1}{2\sqrt{g}}\pi^{i}\pi_{i} +\frac{\sqrt{g}}{4}F^{ij}F_{ij}  +A_0 \mathcal{G}-\partial^i\pi^0 A_i  \, ,
\end{align}
where $\psi$ and $\lambda$ are the respective Lagrange multipliers for the constraints (\ref{ref:Const-2b}) and (\ref{ref:Const-2a}) and where the Hamiltonian contains constraint terms besides the usual energy density $\frac12(\mathbf{E}^2 + \mathbf{B}^2)$, which we have included  following \cite{Henneaux:2018hdj,Henneaux:2018gfi} for later convenience.  We have set
\begin{align}
\tilde{\epsilon} & =\frac{\ln r}{r}\tilde{\epsilon}_{\log}^{(1)}+\frac{1}{r}\tilde{\epsilon}^{(1)}+o\left(r^{-1}\right)\,,\\
\tilde{\mu} & =\frac{\ln r}{r^{2}}\tilde{\mu}_{\log}^{(1)}+\frac{1}{r^{2}}\tilde{\mu}^{(1)}+o\left(r^{-2}\right)\,,
\end{align}
with 
\begin{align}
\tilde{\mu}_{\log}^{(1)}-\tilde{\mu}^{(1)} & =\xbar D_{A} \xbar A^{A}+3\xbar A_{r} \,,\label{eq:TildeMu}\\
\tilde{\epsilon}_{\log}^{(1)}-\tilde{\epsilon}^{(1)} & =\xbar\Psi\,.\label{eq:TildeEpsilon}
\end{align}
Here, the functions of the angles $\xbar A^{A}$, $\xbar A_{r}$ and $\xbar\Psi$ are the $\mathcal O(1)$ coefficients appearing in the asymptotic expansion of the vector potential $A_\mu$, see (\ref{eq:DecayAr})-(\ref{eq:DecayA0}) below.  The functions $\tilde{\mu}_{\log}^{(1)}$, $\tilde{\mu}^{(1)}$, 
$\tilde{\epsilon}_{\log}^{(1)}$ and $\tilde{\epsilon}^{(1)}$ are not completely determined by the equations (\ref{eq:TildeMu}) and (\ref{eq:TildeEpsilon}) but only the combinations $\tilde{\mu}_{\log}^{(1)}-\tilde{\mu}^{(1)}$ and $\tilde{\epsilon}_{\log}^{(1)}-\tilde{\epsilon}^{(1)}$ are physically relevant (see below).   With the inclusion of the constraint terms, the Hamiltonian density $\mathcal{H} + \tilde \epsilon \mathcal G + \tilde \mu \pi^0$ coincides with the density of the Poincar\'e generator of time translations discussed below. 
Finally, the term $\mathfrak{B}$ is the integral over time of a surface term which we will write explicitly once we have given the boundary conditions.

The constraint function $\mathcal{G}$ generates the gauge transformation 
\be \delta A_i = \partial_i \epsilon\, , \label{eq:Gauge-5a}
\ee
 while the other constraint function $\pi^0$ generates the gauge transformations 
 \be
 \delta A_0 = \mu \, .  \label{eq:Gauge-5b}
 \ee
 These are proper in the sense of \cite{Benguria:1976in} if $\epsilon$ and $\mu$ decrease sufficiently fast at infinity, i.e.
 \be
 \epsilon = o\left(r^{-1}\right) \,, \qquad \mu = o\left(r^{-2}\right)\,,   \label{eq:Gauge-5c}
 \ee  as we shall explicitly show in Section \ref{sec:Improper} below.   We will assume that these fall-off conditions are fulfilled until we discuss improper gauge transformations, which can only be meaningfully done after the boundary conditions have been made precise, a task which we have not achieved yet.
 
Since the Lagrange multipliers parametrize the gauge transformation performed in the course of the evolution (on top of the evolution generated by the Hamiltonian), we take for them the same asymptotic decay
 \be
 \psi = o\left(r^{-1}\right) \,, \qquad \lambda = o\left(r^{-2}\right)\, .
 \ee
 Again, more general asymptotic behaviours can be considered once one has full control of the improper gauge symmetries.
 
The action is invariant under the gauge transformations (\ref{eq:Gauge-5a}) and (\ref{eq:Gauge-5b})  provided one transforms at the same time the Lagrange multipliers as
\be
\delta \psi = \dot \epsilon - \mu - \delta \tilde \epsilon\, ,\qquad \delta \lambda = \dot\mu - \triangle \epsilon - \delta \tilde \mu 
\ee
($\delta \tilde \epsilon$ and $\delta \tilde \mu$ preserve the asymptotic decays of $\psi$ and $\lambda$ because $\delta \tilde{\mu}_{\log}^{(1)}= \delta \tilde{\mu}^{(1)} =
\delta \tilde{\epsilon}_{\log}^{(1)}= \delta \tilde{\epsilon}^{(1)} = 0$).
The action (\ref{eq:Action-Extended0}) is called the ``extended action'' because the gauge parameters $\epsilon$, $\mu$ of the constraints are taken to be independent \cite{Dirac1967,Henneaux:1992ig}.  This is the form that  exhibits most explicitly the symmetry.  The formulation that emerges from the Maxwell Lagrangian is characterized by $\psi = - \tilde \epsilon$ (a condition that is permissible once one allows for more general asymptotic behaviours of the Lagrange multipliers), which relates $\epsilon$ and $\mu$ through $\mu = \dot{\epsilon}$ so that $\delta A_\mu = \partial_\mu \epsilon$. The two formulations are physically equivalent because $\dot \epsilon$ and $\epsilon$ are independent at any given time \cite{Dirac1967,Henneaux:1992ig}.  %We shall come back to this question in the concluding section.

\subsection{Asymptotic conditions}
The gauge transformations  (\ref{eq:Gauge-5a}), (\ref{eq:Gauge-5b}) and (\ref{eq:Gauge-5c})  are abelian so that their finite form coincides with their infinitesimal ones.  The boundary conditions on the canonical variables are taken to differ from those of \cite{Henneaux:2018hdj,Henneaux:2018gfi} by gauge transformation terms $\Delta A_i = \partial_i \Theta$, $\Delta A_0 = \Xi$ with finite gauge parameter ($\Theta$, $\Xi $) that contain contributions that are of order $\mathcal{O}(r)$ and $\mathcal{O}(\ln r)$ with respect to the leading orders present in \cite{Henneaux:2018hdj,Henneaux:2018gfi}, i.e.
$$ \Theta =  r \Phi_{\text{lin}} + \ln r\,\Phi_{\log} + \Phi + \frac{\ln r}{r}\Phi_{\log}^{(1)} +o\left(\frac{\ln r}{r}\right),$$
$$ \Xi = \Psi_{\text{lin}}+\frac{\ln r}{r}\Psi_{\log}+\frac{1}{r}\xbar\Psi+\frac{\ln r}{r^{2}}\Psi_{\log}^{(1)} + o\left(\frac{\ln r}{r^2}\right).$$
The terms $\Phi_{\text{lin}}$, $\Phi_{\log}$, $\Phi^{(1)}_{\log}$, $\Psi_{\text{lin}}$, $\Psi^{(1)}_{\log}$ and $\Psi_{\log}$ are absent in \cite{Henneaux:2018hdj,Henneaux:2018gfi}.

These gauge transformations only affect the components of the vector potential since the momenta are gauge invariant. In polar coordinates,
\be
ds^2 = - dt^2 + dr^2 +g_{AB} dx^A dx^B\,, \qquad {g_{AB}} = r^2 \xbar g_{AB}\,,
\ee
where $\xbar g_{AB}$ ($A, B = 1,2$) is the round metric of the unit sphere, the asymptotic conditions read explicitly 
\begin{align}
A_{r} & =\Phi_{\text{lin}}+\frac{1}{r}\xbar A_{r}+\frac{\ln r}{r^{2}}A_{r}^{\log(2)}+\frac{1}{r^{2}}A_{r}^{(2)}+o\left(r^{-2}\right)\,, \label{eq:DecayAr} \\
A_{A} & =r\partial_{A}\Phi_{\text{lin}}+\ln r\,\partial_{A}\Phi_{\log}+\xbar A_{A}+\frac{\ln r}{r}A_{A}^{\log(2)}+\frac{1}{r}A_{A}^{(2)}+o\left(r^{-2}\right)\,,\label{eq:DecayAA} \\
& \quad A_{r}^{\log(2)}=-\Phi_{\log}^{(1)}, \quad A_{A}^{\log(2)}=\partial_{A}\left(\Phi_{\log}^{(1)}\right) \quad \Leftrightarrow \quad \partial_A A_{r}^{\log(2)} + A_{A}^{\log(2)} = 0 \,, \label{eq:BCSubleading}\\
A_0 & =\Psi_{\text{lin}}+\frac{\ln r}{r}\Psi_{\log}+\frac{1}{r}\xbar\Psi+\frac{\ln r}{r^{2}}\Psi_{\log}^{(1)}+\frac{1}{r^{2}}\Psi^{(1)}+o\left(r^{-2}\right)\,,\label{eq:DecayA0}
\end{align}
for the vector potential
and 
\begin{align}
\pi^{r} & =\xbar\pi^{r}+\frac{1}{r}\pi_{(2)}^{r}+\frac{1}{r^{2}}\pi_{(3)}^{r}+\mathcal{O}\left(r^{-3}\right)\,,\\
\pi^{A} & =\frac{1}{r}\xbar\pi^{A}+\frac{1}{r^{2}}\pi_{(2)}^{A}+\frac{1}{r^{3}}\pi_{(3)}^{A}+\mathcal{O}\left(r^{-4}\right)\, , \\
\pi^0 & =\frac{1}{r^{2}}\pi_{\Psi}^{(2)}+o\left(r^{-2}\right)\,, \label{eq:DecayPi0}
\end{align}
for the momenta, which carry density weight one.  Here, the coefficients  in the expansion of $r$ and $\ln r$ depend only on the angles and are subject to the following parity conditions,
\begin{align}
\xbar A_{r} & =\text{odd}\,,\quad\xbar\pi^{r}=\text{even}\,,\quad\xbar A_{A}=(\xbar A)_{\text{even}}+\partial_{A}\Phi\, \; \; (\Phi=\text{even})\,,\quad \xbar\pi^{A}=\text{odd}\,, \label{eq:Parity0}\\
\Phi_{\text{lin}} & =\text{odd}\,,\quad\Phi_{\log}=\text{odd}\,,\quad\Psi_{\text{lin}}=\text{even}\,,\quad\Psi_{\text{\ensuremath{\log}}}=\text{even}\,.\label{eq:Parity1}
\end{align}
The parity conditions (\ref{eq:Parity0}) reduce to those of \cite{Henneaux:2018hdj,Henneaux:2018gfi}.  We will mention as we proceed where the parity conditions (\ref{eq:Parity1}) on the new terms are needed.  The parity conditions (\ref{eq:Parity1}) are simply enforced by allowing only $\Theta$ and $\Xi$ in $\Delta A_i = \partial_i \Theta$, $\Delta A_0 = \Xi$ that fulfill these parity conditions. Note that if $\Phi$ had an odd part, it could be absorbed in $(\xbar A)_{\text{even}}$ so that we can assume that $\Phi$ is even.   

We have kept the terms of order $\mathcal{O}(r^{-2})$ in $A_r$ and $\mathcal{O}(r^{-1})$ in $A_A$ because they now contribute to the charges.  Similarly, $\pi_{(2)}^{r}$  becomes  physically relevant. Note that the intermediate terms $A_{r}^{\log(2)}$ and $A_{A}^{\log(2)}$, absent in \cite{Henneaux:2018hdj,Henneaux:2018gfi}, have  their entire origin 
in the gauge transformation with  gauge parameter $\epsilon=\frac{\ln r}{r}\Phi_{\log}^{(1)}$, hence (\ref{eq:BCSubleading}).   

Once the boundary conditions have been specified, one can write down explicitly the boundary terms in the action.  The constructive procedure that we have followed for doing so is explained in Appendix {\bf \ref{AppendixA}}. It leads to
\be
\mathfrak{B} = \int dt \oint_{S_2^\infty}  d^{2}x\mathcal{S}   - \int dt \oint_{S_2^\infty} d^{2}x\mathcal{B} \,,\label{eq:Boundary-Action-Extended} \\
\ee
where  $\oint d^{2}x\mathcal{S}$ is a surface integral which is linear in the time derivatives of the canonical variables,
\be
\mathcal{S} = \sqrt{\xbar g}\Big(-\xbar A_{r}\dot{\xbar\Psi}+\Psi_{\log}\dot{\Phi}-A_{r}^{(2)}\dot{\Psi}_{\text{lin}}+\xbar\Psi^{(1)}\dot{\Phi}_{\text{lin}}\Big)\,, \qquad \xbar\Psi^{(1)}\equiv\Psi_{\log}^{(1)}-\Psi^{(1)}\,, \label{eq:DefS}
\ee
while $\mathcal B$ does not depend on the time derivatives of the canonical variables and is given by
\be
\mathcal B = \xbar\Pi^{r}\Psi_{\text{lin}}+\sqrt{\xbar g}\,\partial_{A}\xbar A_{r}\xbar D^{A}\Phi_{\text{lin}}-2\sqrt{\xbar g}\,\xbar A_{r}\Phi_{\text{lin}}\,,
\ee
with
\be
\xbar\Pi^{r}  =\xbar\pi^{r}+\sqrt{\xbar g}\,\Psi_{\log}\,.
\ee
It should be noted that the zero mode of $\Phi$, which drops from the potential $A_i$, is present in the action through the boundary kinetic term.  Its conjugate is non trivial even in the absence of charged fields, and given by the zero mode of $\Psi_{\log}$. 

The boundary term $\oint d^{2}x\mathcal{S}$ completes the  boundary term introduced in \cite{Henneaux:2018gfi} in a way that integrability of the boost charges, which was the very reason for introducing it there,  is maintained with our more general boundary conditions.  Because $\mathcal S$ involves the time derivatives of the fields, it contributes to the symplectic form, which reads explicitly 
\begin{align}
\Omega & =\int d^{3}x\big(d_{V}\pi^{i}d_{V}A_{i}+d_{V}\pi^0 d_{V}A_0\big) \nonumber\\
 & \quad-\oint d^{2}x\sqrt{\xbar g}\left[d_{V}\xbar A_{r}d_{V}\xbar\Psi-d_{V}\Psi_{\log}d_{V}\Phi+d_{V}A_{r}^{(2)}d_{V}\Psi_{\text{lin}}-d_{V}\xbar\Psi^{(1)}d_{V}\Phi_{\text{lin}}\right]\,.
\end{align}
This symplectic form is non-degenerate in the sense that if $X$ is a phase space vector field such that $\iota_X \Omega = 0$, then $X=0$.

The symplectic form pairs $\xbar A_r$ with $\xbar \Psi$, as in  \cite{Henneaux:2018gfi}, and also introduces a surface conjugate $\Psi_{\log}$ to $\Phi$, which had none with the earlier boundary conditions since there was no logarithmic term $\ln r /r$ in $A_0$. The new variables $\Psi_{\text{lin}}$ and $\Phi_{\text{lin}}$, which would be absent if we had limited the extension of the gauge transformations of  \cite{Henneaux:2018gfi} to gauge transformations blowing only logarithmically at infinity (with no linear term),  are naturally paired with the subleading terms in the expansion of the components of the potential. This canonical structure will be reflected in the Poisson brackets of the improper gauge charges. In fact, the brackets of these charges  conversely restrict  the symplectic structure to the above form, which provides an independent argument (besides integrability of the boost charges) for extending the $\mathcal S$ of \cite{Henneaux:2018gfi}  as in (\ref{eq:DefS}).

As a side final comment to this subsection, we note that the pair $(A_0, \pi^0)$ was denoted $(\Psi, \pi_\Psi)$ in \cite{Henneaux:2018hdj,Henneaux:2018gfi}, and that the Lagrange multiplier $\psi$ was denoted $A_0$.  This was due to the peculiar constructive way they were arrived at by studying the dynamics at the boundary. We revert here to more familiar notations.

\subsection{Asymptotic form of the constraints}

We shall also require that the constraints decay one order faster than the one prescribed in \cite{Henneaux:2018gfi}.  This yields (\ref{eq:DecayPi0}) for $\pi^0$ as well as $\mathcal{G} = \mathcal{O}(r^{-3})$, which implies, since
\begin{equation}
\mathcal{G}=\frac{1}{r}\partial_{A}\xbar\pi^{A}+\frac{1}{r^{2}}\left(\partial_{A}\pi_{(2)}^{A}-\pi_{(2)}^{r}\right)+\frac{1}{r^{3}}\left(\partial_{A}\pi_{(3)}^{A}-2\pi_{(3)}^{r}\right)+\mathcal{O}\left(r^{-4}\right)\,,
\end{equation}
the conditions
\begin{equation}
\partial_{A}\xbar\pi^{A}=0\quad\text{and}\quad\partial_{A}\pi_{(2)}^{A}-\pi_{(2)}^{r}=0\,.\label{eq:Constraints}
\end{equation}
These conditions are part of the asymptotic conditions on the canonical variables.

\subsection{Finiteness of the kinetic term}

It is clear that the only potentially divergent term in the action is the bulk kinetic term $\int dtd^{3}x\,\pi^{i}\dot{A}_{i} $.  We check in this subsection that it is finite. One finds by direct substitution 
\begin{align}
\int dt\,dr\,d^{2}x\,\pi^{i}\dot{A}_{i} & =\int dt\,dr\,d^{2}x\Big[\xbar\pi^{r}\dot{\Phi}_{\text{lin}}+\xbar\pi^{A}\partial_{A}\dot{\Phi}_{\text{lin}}+\frac{\ln r}{r}\xbar\pi^{A}\partial_{A}\dot{\Phi}_{\log}\\
 & \quad+\frac{1}{r}\left(\xbar\pi^{A}\dot{\xbar A}_{A}+\xbar\pi^{r}\dot{\xbar A}_{r}+\pi_{(2)}^{r}\dot{\Phi}_{\text{lin}}+\pi_{(2)}^{A}\partial_{A}\dot{\Phi}_{\text{lin}}\right)+\mathcal{O}\left(r^{-2}\right)\Big]\,.
\end{align}

i) For the linear divergence, we note immediately that the first term
is zero because $\Phi_{\text{lin}}$ is \emph{odd}, while the second
vanishes by virtue of the condition $\partial_{A}\xbar\pi^{A}=0$.
\newline

ii) The term proportional to $r^{-1}\ln r$ vanishes by virtue of
$\partial_{A}\xbar\pi^{A}=0$. \newline

iii) The terms proportional to the logarithmic divergence reduce to
(after considering parity conditions)
\begin{equation}
\oint d^{2}x\left[-\partial_{A}\xbar\pi^{A}\dot{\Phi}-\left(\partial_{A}\pi_{(2)}^{A}-\pi_{(2)}^{r}\right)\dot{\Phi}_{\text{lin}}\right]\,,
\end{equation}
which vanish by virtue of the conditions in \eqref{eq:Constraints}.
\newline

The bulk kinetic term and hence the action are finite.  The parity conditions on $\Phi_{\text{lin}}$ and on the asymptotic fields appearing already in \cite{Henneaux:2018hdj,Henneaux:2018gfi}, as well as the fast asymptotic decay of the constraint functions, are key for this result.

\subsection{Equations of motion}
\label{Sub:EOM}

The equations of motion that follow from the action can be analyzed as follows.
\begin{itemize}
\item Variation with respect to the Lagrange multipliers yields the constraints $\mathcal G \approx 0$ and $\pi^0 \approx 0$.
\item Variation with respect to the field $A_0$ yields, besides the bulk equation of motion  $\dot{\pi}^0 + \mathcal{G} = 0$ which is a consequence of the constraints,  the following conditions on the asymptotic fields,
\begin{align}
\dot{\xbar A}_{r} & =\dot{\Phi}_{\text{lin}}=0\,, \label{eq:DotAr}\\
\dot{A}_{r}^{(2)} & =\frac{1}{\sqrt{\xbar g}}\left(\xbar\pi^{r}+\sqrt{\xbar g}\,\Psi_{\log}\right)\,,\label{eq:pir-1}\\
\dot{\Phi} & =\Psi_{\text{lin}}\,. \label{eq:DotPhi}
\end{align}
\item Variation with respect to $\pi^0$ yields the bulk equation of motion
\be
\dot{A}_0 - \lambda - \partial^iA_i - \tilde \mu= 0  \label{eq:EOMA0}
\ee
(without surface term contribution because $\pi^0$ decreases sufficiently fast at infinity).   This equation implies 
\begin{align}
\dot{\Psi}_{\text{lin}} & =0\,,\\
\dot{\Psi}_{\log} & =0\,,\\
\dot{\xbar\Psi} & =\xbar\triangle\,\Phi_{\text{lin}}+2\Phi_{\text{lin}}\,,\\
\dot{\xbar\Psi}^{(1)} & =\xbar\triangle\,\Phi_{\log}+2\xbar A_{r}\,,
\end{align}
to the leading orders to which $\lambda= o\left(r^{-2}\right)$ does not contribute.  The next orders
can then be used to express the subleading terms in $\dot{A}_0$ to the Lagrange multiplier $\lambda$. 
\item 
Variation of the action with respect to the conjugate momenta $\pi^{i}$
yields the equation 
\be
\dot{A}_{i}-\partial_{i}\left(A_{0}+\psi+\tilde{\epsilon}\right)-\frac{\pi_{i}}{\sqrt{g}} = 0 \, , 
\ee
or equivalently
\begin{equation}
\pi_{i}=\sqrt{ g}\left(\partial_{t}A_{i}-\partial_{i}\left(A_{0}+\psi+\tilde{\epsilon}\right)\right)\,.
\end{equation}
The asymptotic form of the field equations yields the following conditions
\begin{align}
\dot{\Phi}_{\text{lin}} & =0\,,\\
\dot{\xbar A}_{r} & =0\,,\\
\dot{A}_{r}^{\log(2)}+\Psi_{\log} & =0\,,\\
\xbar\pi^{r} & =\sqrt{\xbar g}\left(\dot{A}_{r}^{(2)}-\Psi_{\log}\right)\,,\label{eq:pir-2}\\
\pi_{(2)}^{r} & =\sqrt{\xbar g}\left(\dot{A}_{r}^{(3)}+2\Psi^{(1)}-\Psi^{(1)}_{\log}+2\psi^{(2)}-\psi^{(2)}_{\log}+2\tilde{\epsilon}^{(2)}-\tilde{\epsilon}^{(2)}_{\log}\right)\,,
\end{align}
which are consistent with the equations (\ref{eq:DotAr}) and (\ref{eq:pir-1}) obtained previously.
\item Variation of the action with respect to the gauge
field $A_{i}$ yields for the bulk term
 (up to a total time derivative):
\begin{equation}
\delta I_{\text{bulk}}=\int dt\Big[-\int d^{3}x\Big(\dot{\pi}^{i}+\sqrt{g}\nabla_{j}F^{ij}-\partial^{i}\pi^0\Big)\delta A_{i}-\oint d^{2}x\sqrt{\xbar g}\,\xbar\triangle\,\Big(\xbar A_{r}-\Phi_{\log}\Big)\delta\Phi_{\text{lin}}\Big]\,,
\end{equation}
and for the boundary term
\begin{align}
\delta I_{\text{boundary}}=-\int dt\oint d^{2}x\sqrt{\xbar g}\Big\{ & \Big[\dot{\xbar\Psi}-\Big(\xbar\triangle\,\Phi_{\text{lin}}+2\Phi_{\text{lin}}\Big)\Big]\delta\xbar A_{r}+\dot{\Psi}_{\log}\delta\Phi\\
 & +\dot{\Psi}_{\text{lin}}\delta A_{r}^{(2)}+\Big[\dot{\xbar\Psi}^{(1)}-\Big(\xbar\triangle\,\xbar A_{r}+2\xbar A_{r}\Big)\Big]\delta\Phi_{\text{lin}}\Big\}\,.
\end{align}
This implies the equations
\begin{align}
\dot{\pi}^{i} & =-\sqrt{g}\nabla_{j}F^{ij}+\partial^{i}\pi^0\,,\\
\dot{\xbar\Psi}^{(1)} & =-\xbar\triangle\,\Big(\xbar A_{r}-\Phi_{\log}\Big)+\xbar\triangle\,\xbar A_{r}+2\xbar A_{r}=\xbar\triangle\,\Phi_{\log}+2\xbar A_{r}\,,\\
\dot{\xbar\Psi} & =\xbar\triangle\,\Phi_{\text{lin}}+2\Phi_{\text{lin}}\,,\\
\dot{\Psi}_{\log} & =\dot{\Psi}_{\text{lin}}=0\,,
\end{align}
which are consistent with the dynamical equations for the asymptotic components of $A_0$ derived above.

The time derivatives of the first three leading orders of the conjugate
momentum are then given by
\begin{align}
\dot{\xbar\pi}^{r} & =0\,,\\
\dot{\pi}_{(2)}^{r} & =\sqrt{\xbar g}\,\xbar\triangle\,\Big(\xbar A_{r}-\Phi_{\log}\Big)\,,\\
\dot{\pi}_{(3)}^{r} & =\sqrt{\xbar g}\,\Big[\xbar\triangle\,A_{r}^{(2)}+\xbar D_{A}\Big(A^{(2)A}-A^{\log(2)A}\Big)\Big]\,.
\end{align}
Note that while the leading ``monopole'' term $\xbar\pi^r$ is conserved, the subsequent terms are not.  This is of course well known, but will not prevent us from defining a conserved quantity involving the subleading term $\pi_{(2)}^{r}$ and generating gauge transformations that blow up at infinity, as achieved in \cite{Seraj:2016jxi} along different lines.

\end{itemize}

This concludes the discussion of the equations of motion.

\section{Improper gauge symmetries}
\label{sec:Improper}

If one allows the parameters $\epsilon$, $\mu$ in (\ref{eq:Gauge-5a}) and (\ref{eq:Gauge-5b}) to decrease slowlier at infinity, in a way compatible with the asymptotic form of the gauge potential,
\begin{align}
\epsilon &=r\epsilon_{\text{lin}}+\ln r\epsilon_{\log}+\xbar\epsilon+\frac{\ln r}{r}\epsilon_{\log}^{(1)}+\frac{1}{r}\epsilon^{(1)}+o\left(r^{-1}\right)\,, \\
\mu & =\mu_{\text{lin}}+\frac{\ln r}{r}\mu_{\log}+\frac{1}{r}\xbar\mu+\frac{\ln r}{r^{2}}\mu_{\log}^{(1)}+\frac{1}{r^{2}}\mu^{(1)}+o\left(r^{-2}\right)\,,
\end{align}
with
\be
\epsilon_{\text{lin}}  =\text{odd}\,,\quad\epsilon_{\log}=\text{odd}\,,\quad \xbar \epsilon = \text{even}\, ,\quad\mu_{\text{lin}}=\text{even}\,,\quad\mu_{\text{\ensuremath{\log}}}=\text{even}\,, \label{eq:ParityGaugePara}
\ee
one finds that the transformations are still symmetries of the action but now of the ``improper gauge type'' \cite{Benguria:1976in}.  The bulk part of the corresponding generator is again given  by the combination  $\int d^{3}x\left(\mu \pi^0+\epsilon\mathcal{G}\right)$ of the constraints, but there are in addition non-vanishing surface terms at infinity.  

The complete generator $Q_X$ of the transformation generated by the phase space vector field $X$ is obtained through the general formula 
\be
\iota_X \Omega = - d_V Q_X \, ,  \label{eq:HamilVF}
\ee
which receives in our case non trivial contributions both from the bulk part of the symplectic form and from its surface part $\mathcal{S}$ \cite{Henneaux:2018gfi}. (If $\Omega$ reduced to its canonical bulk part, the rule (\ref{eq:HamilVF}) would reproduce the integrability condition of \cite{Regge:1974zd}.)

Applying this general rule to the transformations (\ref{eq:Gauge-5a}) and (\ref{eq:Gauge-5b}), which imply the following transformations of the asymptotic fields
\begin{align}
&\delta_{\epsilon}\Phi_{\text{lin}}=\epsilon_{\text{lin}}\,,\quad\delta_{\epsilon}\Phi_{\log}=\epsilon_{\log}\,,\quad\delta_{\epsilon}\Phi=\xbar\epsilon\,,\quad\delta_{\epsilon}\xbar A_{r}=\epsilon_{\log}\,,\quad\delta_{\epsilon}A_{r}^{(2)}=\xbar \epsilon^{(1)} \, , \quad \xbar \epsilon^{(1)} \equiv \epsilon_{\log}^{(1)}-\epsilon^{(1)}\,, \\
&\delta_{\mu}\Psi_{\text{lin}}=\mu_{\text{lin}}\,,\quad\delta_{\mu}\Psi_{\log}=\mu_{\log}\,,\quad\delta_{\mu}\xbar\Psi=\xbar\mu\,,\quad\delta_{\mu}\xbar \Psi^{(1)}=\xbar\mu^{(1)}\, , \quad \xbar\mu^{(1)}\equiv\mu_{\log}^{(1)}-\mu^{(1)} \,,
\end{align}
one finds that the canonical generator of the improper gauge symmetries reads
\begin{equation}
G_{\mu,\epsilon}=\int d^{3}x\left(\mu\pi^0+\epsilon\mathcal{G}\right) + Q_{\epsilon_{\text{lin}}} + Q_{\epsilon_{\log}} + Q_{\xbar\epsilon} + Q_{\xbar \epsilon^{(1)}} + Q_{\mu_{\text{lin}}} + Q_{\mu_{\log}} + Q_{\xbar\mu} + Q_{\xbar\mu^{(1)}}\,,
\end{equation}
with 
\begin{align}
Q_{\epsilon_{\text{lin}}} & =\oint d^{2}x\epsilon_{\text{lin}}\left[\pi_{(2)}^{r}+\sqrt{\xbar g}\,\left(\Psi_{\log}^{(1)}-\Psi^{(1)}\right)\right]\,,\\
Q_{\epsilon_{\log}} & =\oint d^{2}x\sqrt{\xbar g}\,\epsilon_{\log}\xbar\Psi\,,\\
Q_{\xbar\epsilon} & =\oint d^{2}x\xbar\epsilon\left(\xbar\pi^{r}+\sqrt{\xbar g}\,\Psi_{\log}\right)\,,\\
Q_{\xbar \epsilon^{(1)}} & =\oint d^{2}x\sqrt{\xbar g}\,\xbar \epsilon^{(1)}\Psi_{\text{lin}}\,,
\end{align}
and
\begin{align}
Q_{\mu_{\text{lin}}} & =-\oint d^{2}x\sqrt{\xbar g}\,\mu_{\text{lin}}A_{r}^{(2)}\,,\\
Q_{\mu_{\log}} & =-\oint d^{2}x\sqrt{\xbar g}\,\mu_{\log}\,\Phi\,,\\
Q_{\xbar\mu} & =-\oint d^{2}x\sqrt{\xbar g}\,\xbar\mu\,\xbar A_{r}\,,\\
Q_{\xbar\mu^{(1)}} & =-\oint d^{2}x\sqrt{\xbar g}\,\xbar\mu^{(1)}\Phi_{\text{lin}}\,.
\end{align}

In the charge $Q_{\xbar \epsilon^{(1)}}$, only the even part of $\xbar \epsilon^{(1)}$ contributes since the field $\Psi_{\text{lin}}$ is even.  The odd part of $\xbar \epsilon^{(1)}$ drops out and defines a proper gauge symmetry.   To emphasize this point, one sometimes writes $Q_{\xbar \epsilon^{(1)}_{\text{even}}}$.  Therefore, among the charges generating the improper gauge symmetries parametrized by  $\epsilon$, there are two charges characterized by parameters that are even functions on the sphere ($Q_{\xbar\epsilon}$ and $Q_{\xbar \epsilon^{(1)}_{\text{even}}}$), and two charges characterized by odd parameters ($Q_{\epsilon_{\text{lin}}} $ and $Q_{\epsilon_{\log}}$).

Similarly, among the charges generating the improper gauge symmetries parametrized by  $\mu$, there are two charges characterized by parameters that are odd functions on the sphere ($Q_{\xbar \mu_{\text{odd}}}$ and $Q_{\xbar \mu^{(1)}_{\text{odd}}}$), and two charges characterized by even parameters ($Q_{\mu_{\text{lin}}} $ and $Q_{\mu_{\log}}$).  This is because $\xbar A_{r}$ and $\Phi_{\text{lin}}$ are odd funtions on the sphere, so that $\mu_{\text{even}}$ and $\xbar \mu^{(1)}_{\text{even}}$ drop out from the charges and define proper gauge symmetries.

Furthermore,  if  $\epsilon_{\log}^{(1)}$ and $\epsilon^{(1)}$ are such that $\xbar \epsilon^{(1)} = 0$, the corresponding gauge transformation is proper since it has zero charge.  A similar property holds for  $\mu_{\log}^{(1)}$ and $\mu^{(1)}$.

There is thus a total of eight non trivial improper gauge charges, each characterized by a definite parity under the sphere antipodal map.
The brackets among these charges are easily computed.  They are found to form a centrally extended Abelian
algebra with the following non-zero central charges
\begin{align}
\{G_{\epsilon_{\text{lin}}},G_{\xbar\mu^{(1)}}\} & =-\{G_{\xbar\mu^{(1)}},G_{\epsilon_{\text{lin}}}\}=\oint d^{2}x\sqrt{\xbar g}\,\epsilon_{\text{lin}}\,\xbar\mu^{(1)}\,, \label{eq:Central1}\\
\{G_{\epsilon_{\log}},G_{\xbar\mu}\} & =-\{G_{\xbar\mu},G_{\epsilon_{\log}}\}=\oint d^{2}x \sqrt{\xbar g}\,\epsilon_{\log}\,\xbar\mu\,,\label{eq:Central2}\\
\{G_{\xbar\epsilon},G_{\mu_{\log}}\} & =-\{G_{\mu_{\log}},G_{\xbar\epsilon}\}=\oint d^{2}x\sqrt{\xbar g}\,\xbar\epsilon\,\mu_{\log}\,,\label{eq:Central3}\\
\{G_{\xbar \epsilon^{(1)}},G_{\mu_{\text{lin}}}\} & =-\{G_{\mu_{\text{lin}}},G_{\xbar \epsilon^{(1)}}\}=\oint d^{2}x\sqrt{\xbar g}\, \xbar \epsilon^{(1)}\,\mu_{\text{lin}}\,.\label{eq:Central4}
\end{align}

In the paper \cite{Henneaux:2018gfi} where the boundary conditions are more restrictive, only the improper gauge charges $G_{\xbar\epsilon}$ and $G_{\xbar\mu}$ are present.  These commute, $\{G_{\xbar\epsilon}, G_{\xbar\mu}\} = 0$ so that the central terms are absent.   The improper gauge charges associated with logarithmic gauge transformations ($G_{\epsilon_{\log}}$ and $G_{\mu_{\log}}$) are conjugate to these charges.  The improper gauge transformations linear in $r$ bring in two additional new charges ($G_{\epsilon_{\text{lin}}}$ and $G_{\mu_{\text{lin}}}$) as expected, but at the same time they turn on the subleading gauge transformations, which become non-trivial and also bring in two additional non trivial charges.  These form canonically conjugate pairs.     It is because of the central terms present in the extended improper gauge algebra that one can make the improper gauge generators commute with the Poincar\'e generators in the asymptotic symmetry algebra, as in the gravity case \cite{Fuentealba:2022xsz}, and as we shall show explicitly below.

Among the improper gauge charges, $G_{\xbar\epsilon}$ and $G_{\xbar\mu}$ combine to form the angle-dependent $u(1)$ charge seen at null infinity  \cite{Henneaux:2018gfi}.  The charge $G_{\epsilon_{\text{lin}}}$ involves the $1/r^3$ component of the electric field with an odd gauge parameter and is related to the electric dipole moment for the $\ell = 1$ spherical harmonic.  Together with its companion $G_{\mu_{\text{lin}}}$, we expect that it should be connected to the charges that underlie the subleading soft photon theorems \cite{Lysov:2014csa,Campiglia:2016hvg,Conde:2016csj}. 

Note that the integrands of the charges $Q_{\epsilon_{\text{lin}}} $ and $Q_{\epsilon_{\log}}$ are not conserved in time since it follows from the equations of motion that 
\begin{equation}
\partial_{t}\left(\pi_{(2)}^{r}+\sqrt{\xbar g}\,\xbar\Psi^{(1)}\right)=\xbar\triangle\,\xbar A_{r}+2\xbar A_{r}\,, \qquad \partial_{t} \xbar \Psi  =\xbar\triangle\,\Phi_{\text{lin}}+2\Phi_{\text{lin}} \, .
\end{equation}
This expresses the fact that $Q_{\epsilon_{\text{lin}}} $ and $Q_{\epsilon_{\log}}$ do not commute with the generator $H$ of time translations and hence are conserved only at the price of an explicit time dependence.  Indeed the Poisson brackets with the Poincar\'e generators (prior to decoupling) imply, as we shall see,
\be
\{G_{\epsilon_{\text{lin}}}, H \} = G_{\xbar \mu} \, , \qquad \xbar \mu =-( \xbar\triangle\,\epsilon_{\text{lin}}+2\epsilon_{\text{lin}})
\ee
and 
\be
\{G_{\epsilon_{\log}}, H \} = G_{\xbar  \mu^{(1)}} \, , \qquad \xbar \mu^{(1)} =-( \xbar\triangle\,\epsilon_{\log}+2\epsilon_{\log})\, .
\ee
Similarly,
\be
\{G_{\mu_{\text{lin}}}, H \} = G_{\xbar \epsilon} \, , \qquad \xbar \epsilon = -\mu_{\text{lin}}
\ee
and 
\be
\{G_{\mu_{\log}}, H \} = G_{\xbar  \epsilon^{(1)}} \, , \qquad \xbar \epsilon^{(1)} = - \mu_{\log}\, ,
\ee
in agreement with  (\ref{eq:pir-1}) and (\ref{eq:DotPhi}).

The equations of motion have been derived in Subsection {\bf \ref{Sub:EOM}} assuming that the Lagrange multipliers $\psi$ and $\lambda$ decrease sufficiently fast at infinity so that they define proper gauge symmetries.  The surface term in the action was adjusted under this assumption.  In fact, if we had not known that $\psi = o(r^{-1})$ and $\lambda = o(r^{-2})$, we would have derived from the extremization of the action (\ref{eq:Action-Extended0}) {\it (with that boundary term)} that consistency of the equations of motion implies that $\psi$ and $\lambda$ should define proper gauge transformations ($\psi_{\text{lin}} = 0 = \psi_{\log} = \xbar\psi =  \xbar \psi^{(1)}$ and $\lambda_{\text{lin}} = 0 = \lambda_{\log} = \xbar\lambda =  \xbar \lambda^{(1)}$).  

A more general behaviour of the  Lagrange multipliers $\psi$ and $\lambda$ can be accomodated.  One can allow them to define improper gauge symmetries provided one includes in the action the corresponding surface integral.  This would be in particular needed if one wanted to impose the Lorenz gauge $\dot{A}_0 = \partial^i A_i$ which requires from (\ref{eq:EOMA0}) that $\lambda$ should be equal to $- \tilde{\mu}$.

\section{Poincar\'e invariance}
\label{sec:PoincInv}

\subsection{Poincar\'e transformations}

The analysis of Poincar\'e invariance proceeds along the lines of  \cite{Henneaux:2018gfi}, adapted to the less stringent boundary conditions considered in this paper.   We shall therefore give only the final results and check their consistency.  A more constructive approach, which was actually the approach we followed to arrive at the results presented below,  is given in Appendix {\bf \ref{AppendixA}}.

The $10$-dimensional Poincar\'e symmetry is generated by the vector fields conveniently parametrized in spherical coordinates as 
\begin{align}
\xi & =br+T\,,\\
\xi^{r} & =W\,,\\
\xi^{A} & =Y^{A}+\frac{1}{r}\xbar D^{A}W\,,
\end{align}
where
\begin{equation}
\xbar D_{A}\xbar D_{B}b+\xbar g_{AB}b=0\,,\quad\xbar D_{A}\xbar D_{B}W+\xbar g_{AB}W=0\,,\quad\xbar D_{A}Y_{B}+\xbar D_{B}Y_{A}=0\,,\quad\partial_{A}T=0\,.
\end{equation}
The boost function $b(x^A)$ and the spatial translation function $W(x^A)$ each belong to the three-dimensional space of the spin-$1$ representation of the rotation group.  The vector field $Y^A$ generates infinitesimal rotations.  It is a Killing vector on the sphere and depends also on three parameters. The last parameter of the Poincar\'e group is given by $T$, which is constant.

The Poincar\'e transformations of the fields are given by
\begin{align}
\delta_{\xi,\xi^{i}}A_{i} & =\frac{\xi\pi_{i}}{\sqrt{g}}+\partial_{i}\left(\xi A_0 \right)+\mathcal{L}_{\xi}A_{i}+\delta_{\epsilon_{(T,W)}}A_{i}\,, \label{eq:Poinc1}\\
\delta_{\xi,\xi^{i}}\pi^{i} & =\sqrt{g}\nabla_{m}\big(F^{mi}\xi\big)+\xi\partial^{i}\pi^0+\mathcal{L}_{\xi}\pi^{i}\,,\\
\delta_{\xi,\xi^{i}}A_0 & =\nabla_{i}\left(\xi A^{i}\right)+\mathcal{L}_{\xi} A_0+\delta_{\mu_{(b,T,W)}}A_0\,,\\
\delta_{\xi,\xi^{i}}\pi^0 & =\xi\partial_{i}\pi^{i}+\mathcal{L}_{\xi}\pi^0\,, \label{eq:Poinc4}
\end{align}
These transformations take the same form as in \cite{Henneaux:2018gfi}, except for the additional correction terms $\delta_{\epsilon_{(T,W)}}A_{i}$ and $\delta_{\mu_{(b,T,W)}}A_0$ that must be added in order to preserve integrability of the Poincar\'e charges (see below).  These transformations are improper gauge symmetries with parameters
\be
\epsilon_{(T,W)} =\frac{\ln r}{r}\epsilon_{\log (T,W)}^{(1)}+\frac{1}{r}\epsilon_{(T,W)}^{(1)}+o\left(r^{-1}\right)\,,
\ee
and 
\be
\mu_{(b,T,W)} = \frac{\ln r}{r^{2}}\mu_{\log (b,T,W)}^{(1)}+\frac{1}{r^{2}}\mu_{(b,T,W)}^{(1)}+o\left(r^{-2}\right)\,,
\ee
with 
\be
\xbar \epsilon_{(T,W)}^{(1)} = T \xbar A_0 + \xbar D^{A}W\xbar A_{A}-W\xbar A_{r}\,,
\ee 
and
\be
\xbar \mu_{(b,T,W)}^{(1)} = 4bA_{r}^{(2)}+T\left(\xbar D_{A}\xbar A^{A}+3\xbar A_{r}\right) -\left(W\xbar\pi^{r}-\partial_{A}W\xbar\pi^{A}\right)-\left[-\partial_{A}W\xbar D^{A}\xbar\Psi+W\left(\Psi_{\log}+\xbar\Psi\right)\right] \, .
\ee

Asymptotically expanding (\ref{eq:Poinc1})-(\ref{eq:Poinc4}) leads to the following transformation laws:
\begin{itemize}
\item For the leading orders:
\begin{align}
\delta_{\xi,\xi^{i}}\Phi_{\text{lin}} & =\mathcal{L}_{Y}\Phi_{\text{lin}}+b\Psi_{\text{lin}}\,,\\
\delta_{\xi,\xi^{i}}\Phi_{\log} & =\mathcal{L}_{Y}\Phi_{\log}+b\Psi_{\log}\,,\\
\delta_{\xi,\xi^{i}}\Psi_{\text{lin}} & =\mathcal{L}_{Y}\Psi_{\text{lin}}+\xbar D_{A}\left(b\xbar D^{A}\Phi_{\text{lin}}\right)+3b\Phi_{\text{lin}}\,,\\
\delta_{\xi,\xi^{i}}\Psi_{\log} & =\mathcal{L}_{Y}\Psi_{\text{\ensuremath{\log}}}+\xbar D_{A}\left(b\xbar D^{A}\Phi_{\log}\right)\,,
\end{align}
\item For the subleading orders:
\begin{align}
\delta_{\xi,\xi^{i}}\xbar A_{r} & =\mathcal{L}_{Y}\xbar A_{r}+\frac{b}{\sqrt{\xbar g}}\xbar\pi_{r}+b\Psi_{\log}\,,\\
\delta_{\xi,\xi^{i}}\xbar A_{A} & =\mathcal{L}_{Y}\xbar A_{A}+\frac{b}{\sqrt{\xbar g}}\xbar\pi_{A}+\partial_{A}\left(W\Phi_{\text{lin}}+\xbar D^{B}W\partial_{B}\Phi_{\text{lin}}+b\xbar\Psi+T\Psi_{\text{lin}}\right)\,,\\
\delta_{\xi,\xi^{i}}\xbar\pi^{r} & =\mathcal{L}_{Y}\xbar\pi^{r}+\sqrt{\xbar g}\,\xbar D_{A}\left[b\xbar D^{A}\left(\xbar A_{r}-\Phi_{\log}\right)\right]\,,\\
\delta_{\xi,\xi^{i}}\xbar\pi^{A} & =\mathcal{L}_{Y}\xbar\pi^{A}-\sqrt{\xbar g}\,\xbar g^{AB}\xbar D^{C}\big(b\xbar F_{BC}\big)\,,
\end{align}
\begin{equation}
\delta_{\xi,\xi^{i}}\xbar\Psi=\mathcal{L}_{Y}\xbar\Psi+\xbar D^{A}W\partial_{A}\Psi_{\text{lin}}+\xbar D_{A}\big(b\xbar A^{A}\big)+2b\xbar A_{r}+T\left(\xbar\triangle\,\Phi_{\text{lin}}+2\Phi_{\text{lin}}\right)\,,
\end{equation}
which implies that
\begin{equation}
\delta_{\xi,\xi^{i}}\Phi=\mathcal{L}_{Y}\Phi+W\Phi_{\text{lin}}+\xbar D^{B}W\partial_{B}\Phi_{\text{lin}}+b\xbar\Psi^{\text{odd}}+T\Psi_{\text{lin}}\,.
\end{equation}
\item For the sub-subleading orders:
\begin{align}
\delta_{\xi,\xi^{i}}A_{r}^{(2)} & =\mathcal{L}_{Y}A_{r}^{(2)}+\xbar D^{A}W\left(\partial_{A}\xbar A_{r}-\xbar A_{A}\right)-W\xbar A_{r}\nonumber \\
 & \quad+\frac{b}{\sqrt{\xbar g}}\pi_{(2)}^{r}+b\left(\Psi^{\log(1)}-\Psi^{(1)}\right)+\frac{T}{\sqrt{\xbar g}}\xbar\pi^{r}+T\left(\Psi_{\log}-\xbar\Psi\right)+\xbar\epsilon_{(T,W)}^{(1)}\,,\\
\delta_{\xi,\xi^{i}}A_{A}^{(2)} & =\mathcal{L}_{Y}A_{A}^{(2)}+\xbar D_{B}\xbar D_{A}W\xbar A^{B}+\xbar D^{B}W\xbar D_{B}\xbar A_{A}+\partial_{A}W\xbar A_{r} \nonumber\\
&\quad +W\partial_{A}\Phi_{\log}+\frac{b}{\sqrt{\xbar g}}\pi_{(2)A}+\frac{T}{\sqrt{\xbar g}}\xbar\pi_{A}\,,\\
\delta_{\xi,\xi^{i}}\pi_{(2)}^{r} & =\mathcal{L}_{Y}\pi_{(2)}^{r}+\xbar D_{A}\left(\xbar D^{A}W\xbar\pi^{r}\right)-\partial_{A}W\xbar\pi^{A}\nonumber \\
 & \quad+\sqrt{\xbar g}\,\xbar D_{A}\big[b\big(\xbar D^{A}A_{r}^{(2)}+A^{(2)A}-A^{\log(2)A}\big)\big]+\sqrt{\xbar g}\,\xbar D_{A}\left[T\xbar D^{A}\left(\xbar A_{r}-\Phi_{\log}\right)\big)\right]\,,\\
\delta_{\xi,\xi^{i}}\pi_{(2)}^{A} & =\mathcal{L}_{Y}\pi_{(2)}^{A}+\xbar D_{B}\left(\xbar D^{B}W\xbar\pi^{A}\right)+\xbar D_{B}\xbar D^{A}W\xbar\pi^{B}+\xbar D^{A}W\xbar\pi^{r}-W\xbar\pi^{A} \nonumber\\
&\quad-\sqrt{\xbar g}\,\xbar g^{AB}\xbar D^{C}\big(bF_{BC}^{(2)}\big)+\sqrt{\xbar g}\,\xbar g^{AB}b\left(\partial_{B}A_{r}^{(2)}+A_{B}^{(2)}\right)\nonumber \\
 & \quad-\sqrt{\xbar g}\,\xbar g^{AB}\xbar D^{C}\big(T\xbar F_{BC}\big)+\sqrt{\xbar g}\,T\xbar D^{A}\xbar A_{r}\,,
\end{align}
\begin{align}
\delta_{\xi,\xi^{i}}\Psi_{\log}^{(1)} & =\mathcal{L}_{Y}\Psi_{\log}^{(1)}+\partial_{A}W\xbar D^{A}\Psi_{\log}-W\Psi_{\log}\nonumber\\
&\quad+\xbar D_{A}\left(bA^{\log(2)A}\right)+ bA_{r}^{\log(2)}+T\xbar\triangle\,\Phi_{\text{log}}+\mu_{\log(b,T,W)}^{(1)}\,,\\
\delta_{\xi,\xi^{i}}\Psi^{(1)} & =\mathcal{L}_{Y}\Psi^{(1)}+\partial_{A}W\xbar D^{A}\xbar\Psi+W\left(\Psi_{\log}-\xbar\Psi\right)\nonumber \\
 & \quad+\xbar D_{A}\left(bA^{(2)A}\right)+b\left(A_{r}^{(2)}+A_{r}^{\log(2)}\right)+T\left(\xbar D_{A}\xbar A^{A}+\xbar A_{r}\right)+\mu_{(b,T,W)}^{(1)}\,,\\
&\delta_{\xi,\xi^{i}}\Phi_{\log}^{(1)}=\mathcal{L}_Y \Phi_{\log}^{(1)}-\partial_A W\xbar D^A\Phi_{\log}+b \Psi^{(1)}+T\Psi_{\log}-\epsilon^{(1)}_{\log(T,W)}\,.
\end{align}
\end{itemize}

\subsection{Poincar\'e generators}

The justification of the above definitions of the Poincar\'e transformations of the fields and of the symplectic structure is that the latter is invariant under the former, 
\be
\mathcal L_{X_{\xi,\xi^{i}}} \Omega = 0  \label{eq:InvOmega}
\ee
with $X_{\xi,\xi^{i}}$ the phase space vector field defined by (\ref{eq:Poinc1})-(\ref{eq:Poinc4}).  The verification of (\ref{eq:InvOmega}) is somewhat cumbersome and involves the following key ingredients:
\begin{itemize}
\item The divergent terms in $\mathcal L_{X_{\xi,\xi^{i}}} \Omega$ are zero thanks to the parity conditions and the equation (\ref{eq:BCSubleading}). In particular, the parity conditions (\ref{eq:Parity1}) on the new terms in the asymptotic expansion of the vector potential are needed.
\item The surface term in the symplectic form has been adjusted so that the remaining terms in $\mathcal L_{X_{\xi,\xi^{i}}} \Omega$, which are finite surface terms (with no bulk contribution),  exactly cancel, taking into account the contributions coming from the correcting gauge transformation terms $\delta_{\epsilon_{(T,W)}}A_{i}$ and $\delta_{\mu_{(b,T,W)}}A_0$. As in \cite{Henneaux:2018gfi}, it is actually the requirement that $\mathcal L_{X_{\xi,\xi^{i}}} \Omega$ should vanish for boosts ($\xi = b r$, $\xi^i = 0$) that fixes the form of the surface terms to be added to the standard canonical bulk part of $\Omega$.
\end{itemize}

Since $\mathcal L_{X_{\xi,\xi^{i}}} \Omega =  d_V \iota_{X_{\xi,\xi^{i}}}\Omega$, one can now compute the Poincar\'e generators through the formula
\begin{equation}
 \iota_{X_{\xi,\xi^{i}}}\Omega=-d_{V}P_{\xi,\xi^{i}}\,.
\end{equation}
One gets
\begin{equation}
P_{\xi,\xi^{i}}=\int d^{3}x\left(\xi\mathcal{H}+\xi^{i}\mathcal{H}_{i}+\epsilon_{(T,W)}\mathcal{G}+\mu_{(b,T,W)}\pi^0\right)+\mathcal{B}_{\xi,\xi^{i}}\,,
\end{equation}
where
\be
\mathcal{H}  =\frac{\sqrt{g}}{4}F^{ij}F_{ij}+\frac{1}{2\sqrt{g}}\pi^{i}\pi_{i}-\partial^{i}\pi^0 A_{i}-\partial_{i}\pi^{i}A_0\,,\\
\ee
(as above) and
\begin{align}
\mathcal{H}_{i} & =F_{ij}\pi^{j}-\partial_{j}\pi^{j}A_{i}+\pi^0\partial_{i}A_0\,,\\
\mathcal{B}_{\xi,\xi^{i}} & =\oint d^{2}x\Big\{ b\Big(\xbar\Pi^{r}\xbar\Psi+\Pi_{(2)}^{r}\Psi_{\text{lin}}+\sqrt{\xbar g}\partial_{A}\xbar A_{r}\xbar A^{A}+\sqrt{\xbar g}\,\partial_{A}A_{r}^{(2)}\xbar D^{A}\Phi_{\text{lin}}-\sqrt{\xbar g}\,A_{r}^{(2)}\Phi_{\text{lin}}\Big)\nonumber\\
 & \quad+Y^{A}\left(\xbar\Pi^{r}\xbar A_{A}+\Pi_{(2)}^{r}\partial_{A}\Phi_{\text{lin}}+\sqrt{\xbar g}\,\xbar\Psi\partial_{A}\xbar A_{r}+\sqrt{\xbar g}\,\Psi_{\text{lin}}\partial_{A}A_{r}^{(2)}\right)\nonumber\\
 & \quad+T\Big(\xbar\Pi^{r}\Psi_{\text{lin}}+\sqrt{\xbar g}\,\partial_{A}\xbar A_{r}\xbar D^{A}\Phi_{\text{lin}}-2\sqrt{\xbar g}\,\xbar A_{r}\Phi_{\text{lin}}\Big)\nonumber\\
 & \quad+W\Big[\Big(\xbar\triangle\,\xbar\Pi^{r}+3\xbar\Pi^{r}\Big)\Phi_{\text{lin}}+\xbar D^{A}\xbar\Pi^{r}\partial_{A}\Phi_{\text{lin}}+\sqrt{\xbar g}\,\xbar A_{r}\xbar\triangle\,\Psi_{\text{lin}}+\sqrt{\xbar g}\,\partial_{A}\xbar A_{r}\xbar D^{A}\Psi_{\text{lin}}\Big]\Big\}\,,
\end{align}
with
\begin{align}
\xbar\Pi^{r} & =\xbar\pi^{r}+\sqrt{\xbar g}\,\Psi_{\log}\,,\\
\Pi_{(2)}^{r} & =\pi_{(2)}^{r}+\sqrt{\xbar g}\,\left(\Psi_{\log}^{(1)}-\Psi^{(1)}\right)\,.
\end{align}

\section{Asymptotic symmetry algebra}
\label{sec:ASA}

A direct computation shows that the algebra of the asymptotic symmetries as defined above is  the semi-direct
sum of the Poincar\'e algebra with the above Abelian set of improper gauge
symmetries, endowed with non-trivial central charges. Indeed, the Poisson
brackets of the generators are given by
\begin{align}
\big\{ P_{\xi_{1},\xi_{1}^{i}},P_{\xi_{2},\xi_{2}^{i}}\big\} & =P_{\hat{\xi},\hat{\xi}^{i}}\,,\\
\big\{ G_{\mu,\epsilon},P_{\xi,\xi^{i}}\big\} & =G_{\hat{\mu},\hat{\epsilon}}\,,\\
\big\{ G_{\mu_{1},\epsilon_{1}},G_{\mu_{2},\epsilon_{2}}\big\} & =C_{\{\mu_{1},\epsilon_{1};\mu_{2},\epsilon_{2}\}}\,,
\end{align}
where
\begin{align}
\hat{\xi} & =\xi_{1}^{i}\partial_{i}\xi_{2}-\xi_{2}^{i}\partial_{i}\xi_{1}\,,\\
\hat{\xi}^{i} & =\xi_{1}^{j}\partial_{j}\xi_{2}^{i}-\xi_{2}^{j}\partial_{j}\xi_{1}^{i}+g^{ij}\left(\xi_{1}\partial_{j}\xi_{2}-\xi_{2}\partial_{j}\xi_{1}\right)\,,
\end{align}
and
\begin{align}
\hat{\mu}_{\text{lin}} & =-Y^{A}\partial_{A}\mu_{\text{lin}}-3b\epsilon_{\text{lin}}-\xbar D_{A}\left(b\xbar D^{A}\epsilon_{\text{lin}}\right)\,,\\
\hat{\epsilon}_{\text{lin}} & =-Y^{A}\partial_{A}\epsilon_{\text{lin}}-b\mu_{\text{lin}}\,,\\
\hat{\mu}_{\log} & =-Y^{A}\partial_{A}\mu_{\log}-\xbar D_{A}\left(b\xbar D^{A}\epsilon_{\log}\right)\,,\\
\hat{\epsilon}_{\log} & =-Y^{A}\partial_{A}\epsilon_{\log}-b\mu_{\log}\,,\\
\hat{\xbar\mu} & =-Y^{A}\partial_{A}\xbar\mu-\xbar D_{A}\left(b\xbar D^{A}\xbar\epsilon\right)-T\left(\xbar\triangle\,\epsilon_{\text{lin}}+2\epsilon_{\text{lin}}\right)-\partial_{A}W\xbar D^{A}\mu_{\text{lin}}\,,\\
\hat{\xbar\epsilon} & =-Y^{A}\partial_{A}\xbar\epsilon-b\xbar\mu-T\mu_{\text{lin}}-W\epsilon_{\text{lin}}-\partial_{A}W\xbar D^{A}\epsilon_{\text{lin}}\,,\\
\hat{\xbar\mu}^{(1)} & =-Y^{A}\partial_{A}\xbar\mu^{(1)}+3b\epsilon^{(1)}+\xbar D_{A}\left(b\xbar D^{A}\epsilon^{(1)}\right)-T\left(\xbar\triangle\,\epsilon_{\log}+2\epsilon_{\log}\right)+3W\mu_{\log}-\partial_{A}W\xbar D^{A}\mu_{\log}\,,\\
\hat{\xbar \epsilon}^{(1)} & =-Y^{A}\partial_{A}\xbar \epsilon^{(1)}-b\xbar\mu^{(1)}-T\mu_{\log}-\partial_{A}\left(\xbar D^{A}W\epsilon_{\log}\right)\,.
\end{align}
in addition to (\ref{eq:Central1})-(\ref{eq:Central4}). 

\section{Algebraic decoupling of the $u(1)$ charges}
\label{sec:ASAu(1)}

The presence of invertible central terms in the algebra of the improper gauge symmetries can be used to redefine the Poincar\'e generators in such a way that the new form of the algebra has a direct sum structure: the improper gauge charges commute  with the new Poincar\'e generators.  The redefinition is non-linear and is obtained by just applying the general formulas derived in \cite{Fuentealba:2022xsz}.

It is in our case explicitly achieved by adding to the Poincar\'e transformations the following field-dependent improper gauge
transformations
\begin{align}
\mu_{\text{lin}}^{(b,Y)} & =-\mathcal{L}_{Y}\Psi_{\text{lin}}-3b\Phi_{\text{lin}}-\xbar D_{A}\left(b\xbar D^{A}\Phi_{\text{lin}}\right)\,,\\
\epsilon_{\text{lin}}^{(b,Y)} & =-\mathcal{L}_{Y}\Phi_{\text{lin}}-b\Psi_{\text{lin}}\,,\\
\mu_{\log}^{(b,Y)} & =-\frac{1}{\sqrt{\xbar g}}\mathcal{L}_{Y}\xbar\Pi^{r}-\xbar D_{A}\left(b\xbar D^{A}\xbar A_{r}\right)\,,\\
\epsilon_{\log}^{(b,Y)} & =-\mathcal{L}_{Y}\xbar A_{r}-\frac{b}{\sqrt{\xbar g}}\xbar\Pi^{r}\,,\\
\xbar\mu^{(b,Y,T,W)} & =-\mathcal{L}_{Y}\xbar\Psi-\xbar D_{A}(b\xbar A^{A})-T\left(\xbar\triangle\,\Phi_{\text{lin}}+2\Phi_{\text{lin}}\right)-\partial_{A}W\xbar D^{A}\Psi_{\text{lin}}\,,\\
\xbar\epsilon^{(b,Y,T,W)} & =-\mathcal{L}_{Y}\Phi-b\xbar\Psi-T\Psi_{\text{lin}}-W\Phi_{\text{lin}}-\partial_{A}W\xbar D^{A}\Phi_{\text{lin}}\,,\\
\xbar\mu_{(b,Y,T,W)}^{(1)} & =-\frac{1}{\sqrt{\xbar g}}\mathcal{L}_{Y}\Pi_{(2)}^{r}-3bA_{r}^{(2)}-\xbar D_{A}\left(b\xbar D^{A}A_{r}^{(2)}\right)\nonumber\\
&\quad-T\left(\xbar\triangle\,\xbar A_{r}+2\xbar A_{r}\right)+\frac{1}{\sqrt{\xbar g}}\left(-3W\xbar\Pi^{r}+\partial_{A}W\xbar D^{A}\xbar\Pi^{r}\right)\,,\\
\xbar \epsilon_{(b,Y,T,W)}^{(1)} & =-\mathcal{L}_{Y}A_{r}^{(2)}-\frac{b}{\sqrt{\xbar g}}\Pi_{(2)}^{r}-\frac{T}{\sqrt{\xbar g}}\xbar\Pi^{r}+2W\xbar A_{r}-\partial_{A}W\xbar D^{A}\xbar A_{r}\,.
\end{align}

The parameters of these transformations depend on the charges themselves and preserve integrability.
The charges associated with these gauge transformations are given by
\begin{align}
Q_{\mu,\epsilon}^{\text{extra}} & =-\oint d^{2}x\Big\{ b\Big(\xbar\Pi^{r}\xbar\Psi+\Pi_{(2)}^{r}\Psi_{\text{lin}}+\sqrt{\xbar g}\partial_{A}\xbar A_{r}\xbar A^{A}+\sqrt{\xbar g}\,\partial_{A}A_{r}^{(2)}\xbar D^{A}\Phi_{\text{lin}}-\sqrt{\xbar g}\,A_{r}^{(2)}\Phi_{\text{lin}}\Big)\nonumber\\
 & \quad+Y^{A}\left(\xbar\Pi^{r}\xbar A_{A}+\Pi_{(2)}^{r}\partial_{A}\Phi_{\text{lin}}+\sqrt{\xbar g}\,\xbar\Psi\partial_{A}\xbar A_{r}+\sqrt{\xbar g}\,\Psi_{\text{lin}}\partial_{A}A_{r}^{(2)}\right)\nonumber\\
 & \quad+T\Big(\xbar\Pi^{r}\Psi_{\text{lin}}+\sqrt{\xbar g}\,\partial_{A}\xbar A_{r}\xbar D^{A}\Phi_{\text{lin}}-2\sqrt{\xbar g}\,\xbar A_{r}\Phi_{\text{lin}}\Big)\nonumber\\
 & \quad+W\Big[\Big(\xbar\triangle\,\xbar\Pi^{r}+3\xbar\Pi^{r}\Big)\Phi_{\text{lin}}+\xbar D^{A}\xbar\Pi^{r}\partial_{A}\Phi_{\text{lin}}+\sqrt{\xbar g}\,\xbar A_{r}\xbar\triangle\,\Psi_{\text{lin}}+\sqrt{\xbar g}\,\partial_{A}\xbar A_{r}\xbar D^{A}\Psi_{\text{lin}}\Big]\Big\}\,.
\end{align}
These boundary terms are supplemented by weakly vanishing
bulk terms with gauge parameters $\epsilon^{\text{extra}}_{(b,Y,T,W)}$ and $\mu^{\text{extra}}_{(b,Y,T,W)}$,
the fall-off of which is determined by the above parameters. 

In fact, the above transformations were derived from these charges, which were constructed according to the general formulas of \cite{Fuentealba:2022xsz}, so that integrabilty was not really an issue.

The new Poincar\'e generators read
\begin{equation}
\tilde{P}_{\xi,\xi^{i}}=P_{\xi,\xi^{i}}+G_{\mu,\epsilon}^{\text{extra}}\,,
\end{equation}
Because
\begin{equation}
\mathcal{B}_{\xi,\xi^{i}}+Q_{\mu,\epsilon}^{\text{extra}}=0\,,
\end{equation}
 the new Poincar\'e generators are pure bulk
\begin{equation}
\tilde{P}_{\xi,\xi^{i}}=\int d^{3}x\left[\xi\mathcal{H}+\xi^{i}\mathcal{H}_{i}+(\epsilon_{(T,W)}+\epsilon^{\text{extra}}_{(b,Y,T,W)})\mathcal{G}+(\mu_{(b,T,W)}+\mu^{\text{extra}}_{(b,Y,T,W)})\pi^0\right]\,,
\end{equation}
without surface term.
One can then check that their brackets with the $u(1)$ gauge charges
are weakly zero:
\begin{equation}
\big\{ G_{\mu,\epsilon},\tilde{P}_{\xi,\xi^{i}}\big\}=0\,.
\end{equation}
This implies in particular that they are conserved with no explicit time dependence.
One can also check that the new Poincar\'e generators satisfies the Poincar\'e
algebra:
\begin{equation}
\big\{\tilde{P}_{\xi_{1},\xi_{1}^{i}},\tilde{P}_{\xi_{2},\xi_{2}^{i}}\big\}=\tilde{P}_{\hat{\xi},\hat{\xi}^{i}}\,.
\end{equation}
The computation is direct.  This result is actually guaranteed to hold by the general argument of \cite{Fuentealba:2022xsz}.

\section{Conclusions}
\label{sec:Conclusions}

In this paper, we have consistently extended the asymptotic symmetries of electromagnetism in four spacetime dimensions by allowing improper gauge transformations parametrized by coefficients that blow up at infinity like $\ln r$ and $r$.  We have also shown that the structure of the algebra can be used to redefine the Poincar\'e generators in such a way that there is no $u(1)$ ambiguity in the definition of the Lorentz generators (angular momentum and boost generators (center of mass)).  Even though parametrized by a function of the angles, the $u(1)$ charges transform in the trivial representation of the redefined Lorentz algebra.
The necessary redefinitions are non-linear and can be performed in a direct manner because we have a well-defined Hamiltonian formulation (see \cite{Fuentealba:2022yqt} for a more complete discussion of this point in the context of Poisson manifolds).  In particular, integrability of the redefined transformations is never an issue.  

Similar results, which go beyond the analysis of \cite{Henneaux:2019yqq} by adopting more flexible boundary conditions, can be established in higher dimensions as we shall show in future work.

One might wonder whether one could continue the construction and include gauge transformations that asymptotically blow up like $r^2$, or $r^3$ etc.  While we have not made a systematic analysis of this question, preliminary investigations lead to a negative conclusion, because there are too many divergences in the formalism that cannot be cancelled by suitable choices of the parity conditions, which can only be of two types (even or odd).  This agrees with the conclusion of \cite{Campiglia:2016hvg} (note that the multipole charges of \cite{Seraj:2016jxi} generically diverge in a time-dependent context, since one might then have higher spherical harmonic terms in, say the $1/r$ order of the fields).

The terms linear in $r$ in the gauge parameters are somewhat analogous to the superrotations or the Diff$(S^2)$ transformations considered in the works \cite{Barnich:2009se,Barnich:2010eb,Campiglia:2014yka,Campiglia:2015yka}. Our study gives some hope that these can be consistently included at spatial infinity, but the non-linear complexity of Einstein theory calls for caution in drawing conclusions that might be premature.  Further work is clearly needed to settle satisfactorily this issue.

Finally, it would be of great interest to investigate how our work is explicitly translated at null infinity.  To achieve that task one should first 
go back to the standard  `non-extended' formalism where the Lagrange multiplier $\psi$ is fixed in terms of the other fields since it is the formulation that is directly connected with the standard Lagrangian formulation.  Work along these lines is in progress.

\vskip 6mm

\noindent \textbf{Note Added.} After this paper was completed, we became aware of the recent preprint \cite{Peraza:2023ivy} where the canonical formulation of gauge transformations that blow up at infinity is considered along different lines.

\section*{Acknowledgments}
We thank Hern\'an Gonz\'alez, Alfredo P\'erez, David Tempo and Ricardo Troncoso for useful discussions. M.H. is grateful to the Universidad Adolfo Ib\'a\~nez for kind hospitality while this work was completed. The research of O.F. was partially supported by a Marina Solvay Fellowship. This work was partially supported by  FNRS-Belgium (conventions FRFC PDRT.1025.14 and IISN 4.4503.15), as well as by funds from the Solvay Family.

\appendix

\section{Poincar\'e invariance}
\label{AppendixA}

In this appendix, we provide details  on how the kinetic term (\ref{eq:DefS}) and the asymptotic form of the transformations of the fields under Poincar\'e transformation were arrived at.

\subsection{Non-integrability of the boosts and  time translations with standard symplectic form}
We start with the familiar description of the canonical formalism of gravity in terms of the spatial components $A_i$ of the vector potential and its conjugate momoentum $\pi^i$ (electric field).  The corresponding symplectic form is pure bulk and reads
\begin{equation}
\Omega=\int d^{3}xd_{V}\pi^{i}d_{V}A_{i}\,.\label{eq:OmegaSmall}
\end{equation}

The change of the symplectic form 
$\Omega$
under the Poincar\'e diffeomorphism $\xi=br+T$ normal to the equal time hyperplanes is given by
\begin{equation}
\mathcal{L}_{\xi}\Omega=\oint d^{2}x\sqrt{\xbar g}\,\left(br+T\right)\xbar g^{AB}d_{V}F_{rA}d_{V}A_{B}\,. 
\end{equation}
Taking into account the fall-off of the mixed radial-angular component
of the curvature
\begin{equation}
F_{rA}=-\frac{1}{r}\left(\partial_{A}\xbar A_{r}-\partial_{A}\Phi_{\log}\right)-\frac{\ln r}{r^{2}}\left(\partial_{A}A_{r}^{\log(2)}+A_{A}^{\log(2)}\right)-\frac{1}{r^{2}}\left(\partial_{A}A_{r}^{(2)}+A_{A}^{(2)}-A_{A}^{\log(2)}\right)+o\left(r^{-2}\right)\,,
\end{equation}
and of the angular component of the gauge field, we obtain the following expression,
\begin{align}
\mathcal{L}_{\xi}\Omega & =-\oint d^{2}x\sqrt{\xbar g}\,\left(br+T\right)\left(\partial_{A}d_{V}\xbar A_{r}-\partial_{A}d_{V}\Phi_{\log}\right)\xbar D^{A}d_{V}\Phi_{\text{lin}} \nonumber \\
 & \quad-\ln r\oint d^{2}x\sqrt{\xbar g}\,b\left(\partial_{A}d_{V}A_{r}^{\log(2)}+d_{V}A_{A}^{\log(2)}\right)\xbar D^{A}d_{V}\Phi_{\text{lin}} \nonumber \\
 & -\ln r\oint d^{2}x\sqrt{\xbar g}\,b\left(\partial_{A}d_{V}\xbar A_{r}-\partial_{A}d_{V}\Phi_{\log}\right)D^{A}d_{V}\Phi_{\log}\nonumber \\
 & \quad-\oint d^{2}x\sqrt{\xbar g}\,b\left(\partial_{A}d_{V}\xbar A_{r}-\partial_{A}d_{V}\Phi_{\log}\right)d_{V}\xbar A^{A} \nonumber \\ 
 &-\oint d^{2}x\sqrt{\xbar g}\,b\left(\partial_{A}d_{V}A_{r}^{(2)}+d_{V}A_{A}^{(2)}-d_{V}A_{A}^{\log(2)}\right)\xbar D^{A}d_{V}\Phi_{\text{lin}}\,,
\end{align}
We now make use of the fact that the components $A_{i}^{\log(2)}$
are pure gauge, then $\partial_{A}A_{r}^{\log(2)}+A_{A}^{\log(2)}=0$,
which takes care of the first logarithmic divergence. The remaining
two divergent terms (in $r$ and $\ln r$) vanish by assuming that
$\Phi_{\log}$ is \emph{odd}. After integration by parts and using
that
\begin{align}
\Phi_{\text{lin}} & =\text{odd}\,,\\
\xbar A_{A} & =(\xbar A_{A})_{\text{even}}+\partial_{A}\Phi,
\end{align}
 the finite terms can be re-written as
\begin{align}
\mathcal{L}_{\xi}\Omega & =\oint d^{2}x\sqrt{\xbar g}\,d_{V}\xbar A_{r}\left[\partial_{A}\left(bd_{V}\xbar A^{A}\right)+T\xbar\triangle\,d_{V}\Phi_{\text{lin}}\right]-\oint d^{2}x\sqrt{\xbar g}\,\partial_{A}\left(b\xbar D^{A}d_{V}\Phi_{\log}\right)d_{V}\Phi\nonumber \\
 & \quad+\oint d^{2}x\sqrt{\xbar g}\,d_{V}A_{r}^{(2)}\partial_{A}\left(b\xbar D^{A}d_{V}\Phi_{\text{lin}}\right)+\oint d^{2}x\sqrt{\xbar g}\,\partial_{A}\left(bd_{V}A^{(2)A}\right)d_{V}\Phi_{\text{lin}}\nonumber \\
 & \quad-\oint d^{2}x\sqrt{\xbar g}\,\left[\partial_{A}\left(bd_{V}A^{\log(2)A}\right)+T\xbar\triangle\,d_{V}\Phi_{\text{log}}\right]d_{V}\Phi_{\text{lin}}\,.\label{eq:old-variation}
\end{align}
These are not zero and hence, with the relaxed boundary conditions for $A_i$, Poincar\'e transformations are not canonical transformations for \eqref{eq:OmegaSmall}.

\subsection{New symplectic form and invariance under boosts and  time translations}
In order to obtain an invariant symplectic form, we include $A_0$ and its conjugate $\pi^0$ with the asymptotic behaviour described in the text and adopt as new symplectic form
\begin{align}
\Omega^{\text{new}} & =\int d^{3}x\big(d_{V}\pi^{i}d_{V}A_{i}+d_{V}\pi^0 d_{V}A_0\big) \nonumber\\
 & \quad-\oint d^{2}x\sqrt{\xbar g}\left[d_{V}\xbar A_{r}d_{V}\xbar\Psi-d_{V}\Psi_{\log}d_{V}\Phi+d_{V}A_{r}^{(2)}d_{V}\Psi_{\text{lin}}-\left(d_{V}\Psi_{\log}^{(1)}-d_{V}\Psi^{(1)}\right)d_{V}\Phi_{\text{lin}}\right]\,,  \label{eq:NewSymplectic}
\end{align}
(with $\pi^0 \approx 0$; $\Omega^{\text{new}}$ is just denoted $\Omega$ in the main text since it is the only one appearing there).
The bulk term $d_{V}\pi^0 d_{V}A_0$ is the standard canonical one, the surface term $d_V \xbar A_{r}d_{V}\xbar\Psi$ has been introduced in \cite{Henneaux:2018gfi} while the idea behind the introduction of the other surface terms is that if one takes 
$\delta_{\xi}\Psi_{\text{lin}}  =\xbar D_{A}\left(b\xbar D^{A}\Phi_{\text{lin}}\right)$, $
\delta_{\xi}\Psi_{\log}  =\xbar D_{A}\left(b\xbar D^{A}\Phi_{\log}\right)$, 
$\delta_{\xi}\xbar\Psi  =\xbar D_{A}\big(b\xbar A^{A}\big)+T\xbar\triangle\,d_{V}\Phi_{\text{lin}}$, 
$\delta_{\xi}\Psi_{\log}^{(1)}  =\xbar D_{A}\left(bA^{\log(2)A}\right)+T\xbar\triangle\,\Phi_{\text{log}}$ and 
$\delta_{\xi}\Psi^{(1)}  =\xbar D_{A}\left(bA^{(2)A}\right)$, one eliminates the non-integrable terms (\ref{eq:old-variation}).
However, this introduces other non-integrable terms (since there are additional variations coming from the new terms), so that the approach is not yet complete.

To get integrable boosts, we proceed as in \cite{Henneaux:2018gfi} and adopt the
 transformation laws of the bulk fields shown there to work, namely,
\begin{align}
\delta_{\xi}A_0 & =\nabla_{i}\left(\xi A^{i}\right)\,,\\
\delta_{\xi}\pi^0 & =\xi\partial_{i}\pi^{i}\,, \\
\delta_{\xi}A_{i} & =\frac{\xi\pi_{i}}{\sqrt{g}}+\partial_{i}\left(\xi\Psi\right)\,,\\
\delta_{\xi}\pi^{i} & =\sqrt{g}\nabla_{m}\big(F^{mi}\xi\big)+\xi\partial^{i}\pi^0\,.
\end{align}
This implies 
the following transformation laws for the asymptotic fields
\begin{align}
\delta_{\xi}\Psi_{\text{lin}} & =\xbar D_{A}\left(b\xbar D^{A}\Phi_{\text{lin}}\right)+3b\Phi_{\text{lin}}\,,\\
\delta_{\xi}\Psi_{\log} & =\xbar D_{A}\left(b\xbar D^{A}\Phi_{\log}\right)\,,\\
\delta_{\xi}\xbar\Psi & =\xbar D_{A}\big(b\xbar A^{A}\big)+2b\xbar A_{r}+T\left(\xbar\triangle\,\Phi_{\text{lin}}+2\Phi_{\text{lin}}\right)\,,\\
\delta_{\xi}\Psi_{\log}^{(1)} & =\xbar D_{A}\left(bA^{\log(2)A}\right)+bA_{r}^{\log(2)}+T\xbar\triangle\,\Phi_{\text{log}}\,,\\
\delta_{\xi}\Psi^{(1)} & =\xbar D_{A}\left(bA^{(2)A}\right)+b\left(A_{r}^{(2)}+A_{r}^{\log(2)}\right)+T\left(\xbar D_{A}\xbar A^{A}+\xbar A_{r}\right)\,.  \\
\delta_{\xi}\Phi_{\text{lin}} & =b\Psi_{\text{lin}}\,,\\
\delta_{\xi}\Phi & =b\xbar\Psi^{\text{odd}}+T\Psi_{\text{lin}}\,,\\
\delta_{\xi}\xbar A_{r} & =\frac{b}{\sqrt{\xbar g}}\xbar\pi_{r}+b\Psi_{\log}\,,\\
\delta_{\xi}A_{r}^{(2)} & =\frac{b}{\sqrt{\xbar g}}\pi_{(2)}^{r}+b\left(\Psi^{\log(1)}-\Psi^{(1)}\right)+\frac{T}{\sqrt{\xbar g}}\xbar\pi^{r}+T\left(\Psi_{\log}-\xbar\Psi\right)\,.
\end{align}
These contain additional terms besides the new ones written below (\ref{eq:NewSymplectic}). 

The change in the bulk part of the symplectic form is found to be
\begin{align}
\mathcal{L}_{\xi}\left(\Omega^{\text{new}}\right)_{\text{bulk}} & =\int d^{2}S_{i}\xi d_{V}F^{ij}d_{V}A_{j}+r\oint d^{2}xbd_{V}\xbar\pi^{r}d_{V}\Psi_{\text{lin}}+\ln r\oint d^{2}xbd_{V}\pi_{(2)}^{r}d_{V}\Psi_{\log} \nonumber\\
 & \quad+\oint d^{2}x\left[b\left(d_{V}\xbar\pi^{r}d_{V}\xbar\Psi+d_{V}\pi_{(2)}^{r}d_{V}\Psi_{\text{\text{lin}}}\right)+Td_{V}\xbar\pi^{r}d_{V}\Psi_{\text{lin}}\right]\,.\label{eq:new-bulk-variation}
\end{align}
In order to eliminate the divergent terms we impose the following
parity conditions
\begin{equation}
\Psi_{\text{lin}}=\text{even}\qquad\text{and}\qquad\Psi_{\log}=\text{even}\,.
\end{equation}
The variation of the boundary term of the symplectic form is given
by
\begin{align}
\mathcal{L}_{\xi}\left(\Omega^{\text{new}}\right)_{\text{boundary}} & =\mathcal{L}_{\xi}\Omega^{\text{previous}}-\oint d^{2}x\left[b\left(d_{V}\xbar\pi^{r}d_{V}\xbar\Psi+d_{V}\pi_{(2)}^{r}d_{V}\Psi_{\text{\text{lin}}}\right)+Td_{V}\xbar\pi^{r}d_{V}\Psi_{\text{lin}}\right]\nonumber\\
 & \text{\ensuremath{\quad-\oint d^{2}x\sqrt{\xbar g}\,\left\{ \left[4bd_{V}A_{r}^{(2)}+Td_{V}\left(\xbar D_{A}\xbar A^{A}+3\xbar A_{r}\right)\right]d_{V}\Phi_{\text{lin}}-Td_{V}\xbar\Psi d_{V}\Psi_{\text{lin}}\right\} }},
\end{align}
where $\mathcal{L}_{\xi}\Omega^{\text{previous}}$ is given in \eqref{eq:old-variation},
and is identically cancelled by the first surface integral in \eqref{eq:new-bulk-variation}.

Putting everything together, one finds that the variation of the new symplectic form reduces to the expression
\begin{equation}
\mathcal{L}_{\xi}\Omega^{\text{new}}=-\oint d^{2}x\sqrt{\xbar g}\,\left\{ \left[4bd_{V}A_{r}^{(2)}+Td_{V}\left(\xbar D_{A}\xbar A^{A}+3\xbar A_{r}\right)\right]d_{V}\Phi_{\text{lin}}-Td_{V}\xbar\Psi d_{V}\Psi_{\text{lin}}\right\} \,.
\end{equation}
These terms can be eliminated by performing the following field dependent
subleading gauge transformations
\begin{align}
\xbar\epsilon_{(T)}^{(1)} & =T\xbar\Psi\,,\\
\xbar\mu_{(b,T)}^{(1)} & =4bA_{r}^{(2)}+T\left(\xbar D_{A}\xbar A^{A}+3\xbar A_{r}\right)\,,\label{eq:Corrective-sub-mu}
\end{align}
where $\xbar\epsilon^{(1)}=\epsilon_{\log}^{(1)}-\epsilon^{(1)}$
and $\xbar\mu^{(1)}=\mu_{\log}^{(1)}-\mu^{(1)}$.
Note  that these correcting terms do not vanish even if the boosts are zero since they contain a contribution involving $T$.

\subsection{Invariance of the new symplectic form under spatial translations}

The change of the bulk term of the symplectic form under
a spatial diffeomorphism is given by
\begin{equation}
\mathcal{L}_{\xi^{i}}\left(\Omega^{\text{new}}\right)_{\text{bulk}}=\int d^{2}S_{i}\xi^{i}\left(d_{V}\pi^{j}d_{V}A_{j}+d_{V}\pi^0d_{V}A_0\right)\,,
\end{equation}
which reduces to the non-vanishing boundary term :
\begin{equation}
\mathcal{L}_{\xi^{i}}\left(\Omega^{\text{new}}\right)_{\text{bulk}}=\oint d^{2}x\left(Wd_{V}\xbar\pi^{r}-\partial_{A}Wd_{V}\xbar\pi^{A}\right)d_{V}\Phi_{\text{lin}}\,.
\end{equation}
for our relaxed asymptotic conditions. 
On the other hand, the variation of the boundary term reads
\begin{eqnarray}
\mathcal{L}_{\xi^{i}}\left(\Omega^{\text{new}}\right)_{\text{boundary}}& =& \oint d^{2}x\sqrt{\xbar g}\Big\{ \left[-\partial_{A}W\xbar D^{A}d_{V}\xbar\Psi+  Wd_{V}\left(\Psi_{\log}+\xbar\Psi\right)\right]d_{V}\Phi_{\text{lin}}  \nonumber \\
&& \qquad +\left(\xbar D^{A}Wd_{V}\xbar A_{A}-Wd_{V}\xbar A_{r}\right)d_{V}\Psi_{\text{lin}}\Big\} \,.
\end{eqnarray}
The sum of these terms can be cancelled by corrective gauge transformations
generated by the subleading parameters
\begin{align}
\xbar\epsilon_{(W)}^{(1)} & =\xbar D^{A}W\xbar A_{A}-W\xbar A_{r}\,,\\
\xbar\mu_{(W)}^{(1)} & =-\left(W\xbar\pi^{r}-\partial_{A}W\xbar\pi^{A}\right)-\left[-\partial_{A}W\xbar D^{A}\xbar\Psi+W\left(\Psi_{\log}+\xbar\Psi\right)\right]\,,
\end{align}
which involve only spatial translations.

This completes the discussion of Poincar\'e invariance of the theory.


\begin{thebibliography}{99}

%\cite{Fuentealba:2022xsz}
\bibitem{Fuentealba:2022xsz}
O.~Fuentealba, M.~Henneaux and C.~Troessaert,
``Logarithmic supertranslations and supertranslation-invariant Lorentz charges,''
[arXiv:2211.10941 [hep-th]].

  %\cite{Bondi:1962px}
\bibitem{Bondi:1962px}
H.~Bondi, M.~G.~J.~van der Burg and A.~W.~K.~Metzner,
``Gravitational waves in general relativity. 7. Waves from axisymmetric isolated systems,''
Proc. Roy. Soc. Lond. A \textbf{269} (1962), 21-52
doi:10.1098/rspa.1962.0161

%\cite{Sachs:1962wk}
\bibitem{Sachs:1962wk}
R.~K.~Sachs,
``Gravitational waves in general relativity. 8. Waves in asymptotically flat space-times,''
Proc. Roy. Soc. Lond. A \textbf{270} (1962), 103-126
doi:10.1098/rspa.1962.0206

 %\cite{Sachs:1962zza}
\bibitem{Sachs:1962zza}
  R.~Sachs,
  ``Asymptotic symmetries in gravitational theory,''
  Phys.\ Rev.\  {\bf 128} (1962) 2851.
  %%CITATION = doi:10.1103/PhysRev.128.2851;%%
   
  %\cite{Bergmann:1961zz}
\bibitem{Bergmann:1961zz}
P.~G.~Bergmann,
``'Gauge-Invariant' Variables in General Relativity,''
Phys. Rev. \textbf{124} (1961), 274-278
doi:10.1103/PhysRev.124.274

%\cite{AshtekarLog85}
\bibitem{AshtekarLog85}
A. Ashtekar, ``Logarithmic Ambiguities in the Description of Spatial Infinity,'' Found. Phys. \textbf{15} (1985), 419

 %\cite{BeigSchmidt82}
  \bibitem{BeigSchmidt82}
  R.~Beig and B.~Schmidt, ``Einstein's equations near spatial infinity,'' Commun. Math. Phys. {\bf 87} (1982) 65.
  
 %\cite{Beig:1983sw}
\bibitem{Beig:1983sw}
  R.~Beig,
  ``Integration Of Einstein's Equations Near Spatial Infinity,''
  Proc.  Royal Soc. A
{\bf 1801} (1984) 295--304.
  
%\cite{Ashtekar:1990gc}
\bibitem{Ashtekar:1990gc}
A.~Ashtekar, L.~Bombelli and O.~Reula,
``The covariant phase space of asymptotically flat gravitational fields," PRINT-90-0318 (Syracuse), published in 
``Mechanics, Analysis and Geometry: 200 Years After Lagrange'', ed. Mauro Francaviglia, North-Holland Delta Series, Elsevier (Amsterdam: 1991)

%\cite{Compere:2011ve}
\bibitem{Compere:2011ve}
G.~Comp\`ere and F.~Dehouck,
``Relaxing the Parity Conditions of Asymptotically Flat Gravity,''
Class. Quant. Grav. \textbf{28} (2011), 245016
[erratum: Class. Quant. Grav. \textbf{30} (2013), 039501]
doi:10.1088/0264-9381/28/24/245016
[arXiv:1106.4045 [hep-th]].

  %\cite{Troessaert:2017jcm}
\bibitem{Troessaert:2017jcm}
C.~Troessaert,
``The BMS4 algebra at spatial infinity,''
Class. Quant. Grav. \textbf{35} (2018) no.7, 074003
doi:10.1088/1361-6382/aaae22
[arXiv:1704.06223 [hep-th]].

   %\cite{Henneaux:2018hdj}
\bibitem{Henneaux:2018hdj}
  M.~Henneaux and C.~Troessaert,
  ``Hamiltonian structure and asymptotic symmetries of the Einstein-Maxwell system at spatial infinity,''
  JHEP {\bf 1807} (2018) 171
  [arXiv:1805.11288 [gr-qc]].
  %%CITATION = doi:10.1007/JHEP07(2018)171;%%     
  
  %\cite{Henneaux:2019yax}
\bibitem{Henneaux:2019yax}
M.~Henneaux and C.~Troessaert, ``The asymptotic structure of gravity at spatial infinity in four spacetime dimensions,'' Proc. Steklov Inst.  Math.  {\bf 309} (2020) 127-149 [arXiv:1904.04495 [hep-th]].    

%\cite{Mirbabayi:2016axw}
\bibitem{Mirbabayi:2016axw}
M.~Mirbabayi and M.~Porrati,
``Dressed Hard States and Black Hole Soft Hair,''
Phys. Rev. Lett. \textbf{117}, no.21, 211301 (2016)
doi:10.1103/PhysRevLett.117.211301
[arXiv:1607.03120 [hep-th]].

  %\cite{Bousso:2017dny}
\bibitem{Bousso:2017dny}
R.~Bousso and M.~Porrati,
``Soft Hair as a Soft Wig,''
Class. Quant. Grav. \textbf{34} (2017) no.20, 204001
doi:10.1088/1361-6382/aa8be2
[arXiv:1706.00436 [hep-th]].  

%\cite{Javadinezhad:2018urv}
\bibitem{Javadinezhad:2018urv}
R.~Javadinezhad, U.~Kol and M.~Porrati,
``Comments on Lorentz Transformations, Dressed Asymptotic States and Hawking Radiation,''
JHEP \textbf{01} (2019), 089
doi:10.1007/JHEP01(2019)089
[arXiv:1808.02987 [hep-th]].

%\cite{Javadinezhad:2022hhl}
\bibitem{Javadinezhad:2022hhl}
R.~Javadinezhad, U.~Kol and M.~Porrati,
``Supertranslation-invariant dressed Lorentz charges,''
JHEP \textbf{04} (2022), 069
doi:10.1007/JHEP04(2022)069
[arXiv:2202.03442 [hep-th]].

%\cite{Chen:2021szm}
\bibitem{Chen:2021szm}
P.~N.~Chen, M.~T.~Wang, Y.~K.~Wang and S.~T.~Yau,
``Supertranslation invariance of angular momentum,''
Adv. Theor. Math. Phys. \textbf{25} (2021) no.3, 777-789
doi:10.4310/ATMP.2021.v25.n3.a4
[arXiv:2102.03235 [gr-qc]].

%\cite{Chen:2021zmu}
\bibitem{Chen:2021zmu}
P.~N.~Chen, J.~Keller, M.~T.~Wang, Y.~K.~Wang and S.~T.~Yau,
``Evolution of Angular Momentum and Center of Mass at Null Infinity,''
Commun. Math. Phys. \textbf{386} (2021) no.1, 551-588
doi:10.1007/s00220-021-04053-7
[arXiv:2102.03221 [gr-qc]].

%\cite{Compere:2021inq}
\bibitem{Compere:2021inq}
G.~Comp\`ere and D.~A.~Nichols,
``Classical and Quantized General-Relativistic Angular Momentum,''
[arXiv:2103.17103 [gr-qc]].
  
%\cite{He:2014cra}
\bibitem{He:2014cra}
  T.~He, P.~Mitra, A.~P.~Porfyriadis and A.~Strominger,
  ``New Symmetries of Massless QED,''
  JHEP {\bf 1410} (2014) 112
  [arXiv:1407.3789 [hep-th]].
  %%CITATION = doi:10.1007/JHEP10(2014)112;%%
  
%\cite{Kapec:2015ena}
\bibitem{Kapec:2015ena}
D.~Kapec, M.~Pate and A.~Strominger,
``New Symmetries of QED,''
Adv. Theor. Math. Phys. \textbf{21} (2017), 1769-1785
doi:10.4310/ATMP.2017.v21.n7.a7
[arXiv:1506.02906 [hep-th]].

%\cite{Campiglia:2015qka}
\bibitem{Campiglia:2015qka}
M.~Campiglia and A.~Laddha,
``Asymptotic symmetries of QED and Weinberg\textquoteright{}s soft photon theorem,''
JHEP \textbf{07} (2015), 115
doi:10.1007/JHEP07(2015)115
[arXiv:1505.05346 [hep-th]].

      %\cite{Campiglia:2017mua}
\bibitem{Campiglia:2017mua}
  M.~Campiglia and R.~Eyheralde,
  ``Asymptotic $U(1)$ charges at spatial infinity,''
  JHEP {\bf 1711} (2017) 168
  [arXiv:1703.07884 [hep-th]].
  %%CITATION = doi:10.1007/JHEP11(2017)168;%%

 %\cite{Strominger:2017zoo}
\bibitem{Strominger:2017zoo}
  A.~Strominger,
  ``Lectures on the Infrared Structure of Gravity and Gauge Theory,''
  arXiv:1703.05448 [hep-th].
  %%CITATION = ARXIV:1703.05448;%%

\bibitem{Henneaux:2018gfi}M.~Henneaux and C.~Troessaert,  ``Asymptotic symmetries of electromagnetism at spatial infinity,'' JHEP \textbf{05}, 137 (2018) doi:10.1007/JHEP05(2018)137 [arXiv:1803.10194 [hep-th]].

  %\cite{Lysov:2014csa}
\bibitem{Lysov:2014csa}
  V.~Lysov, S.~Pasterski and A.~Strominger,
  ``Low's Subleading Soft Theorem as a Symmetry of QED,''
  Phys.\ Rev.\ Lett.\  {\bf 113} (2014) no.11,  111601
  [arXiv:1407.3814 [hep-th]].
  %%CITATION = doi:10.1103/PhysRevLett.113.111601;%%
    
     %\cite{Campiglia:2016hvg}
\bibitem{Campiglia:2016hvg}
  M.~Campiglia and A.~Laddha,
  ``Subleading soft photons and large gauge transformations,''
  JHEP {\bf 1611} (2016) 012
  [arXiv:1605.09677 [hep-th]].
  %%CITATION = doi:10.1007/JHEP11(2016)012;%%
  
  %\cite{Conde:2016csj}
\bibitem{Conde:2016csj}
  E.~Conde and P.~Mao,
  ``Remarks on asymptotic symmetries and the subleading soft photon theorem,''
  Phys.\ Rev.\ D {\bf 95} (2017) no.2,  021701
  [arXiv:1605.09731 [hep-th]].
  %%CITATION = doi:10.1103/PhysRevD.95.021701;%%
  
  %\cite{Barnich:2013sxa}
\bibitem{Barnich:2013sxa}
G.~Barnich and P.~H.~Lambert,
``Einstein-Yang-Mills theory: Asymptotic symmetries,''
Phys. Rev. D \textbf{88} (2013), 103006
doi:10.1103/PhysRevD.88.103006
[arXiv:1310.2698 [hep-th]].
  
  %\cite{Coleman:1967ad}
\bibitem{Coleman:1967ad}
S.~R.~Coleman and J.~Mandula,
``All Possible Symmetries of the S Matrix,''
Phys. Rev. \textbf{159} (1967), 1251-1256
doi:10.1103/PhysRev.159.1251


%\cite{Benguria:1976in}
\bibitem{Benguria:1976in}
R.~Benguria, P.~Cordero and C.~Teitelboim,
``Aspects of the Hamiltonian Dynamics of Interacting Gravitational Gauge and Higgs Fields with Applications to Spherical Symmetry,''
Nucl. Phys. B \textbf{122} (1977), 61-99
doi:10.1016/0550-3213(77)90426-6

\bibitem{Dirac1967}
P.~A.~M.~Dirac, ``Lectures on Quantum Mechanics,'' Yeshiva University, Academic Press (New York: 1967)

\bibitem{Henneaux:1992ig}
M.~Henneaux and C.~Teitelboim,
``Quantization of gauge systems,''
Princeton University Press (Princeton: 1992)

 %\cite{Seraj:2016jxi}
\bibitem{Seraj:2016jxi}
A.~Seraj,
``Multipole charge conservation and implications on electromagnetic radiation,''
JHEP \textbf{06} (2017), 080
doi:10.1007/JHEP06(2017)080
[arXiv:1610.02870 [hep-th]]. 


%\cite{Regge:1974zd}
\bibitem{Regge:1974zd}
T.~Regge and C.~Teitelboim,
``Role of Surface Integrals in the Hamiltonian Formulation of General Relativity,''
Annals Phys. \textbf{88} (1974), 286
doi:10.1016/0003-4916(74)90404-7  
    
 %\cite{Fuentealba:2022yqt}
\bibitem{Fuentealba:2022yqt}
O.~Fuentealba, M.~Henneaux, J.~Matulich and C.~Troessaert,
``Asymptotic structure of the gravitational field in five spacetime dimensions: Hamiltonian analysis,''
JHEP \textbf{07} (2022), 149
doi:10.1007/JHEP07(2022)149
[arXiv:2206.04972 [hep-th]]. 

%\cite{Henneaux:2019yqq}
\bibitem{Henneaux:2019yqq}
M.~Henneaux and C.~Troessaert,
``Asymptotic structure of electromagnetism in higher spacetime dimensions,''
Phys. Rev. D \textbf{99} (2019) no.12, 125006
doi:10.1103/PhysRevD.99.125006
[arXiv:1903.04437 [hep-th]].
  
  %\cite{Barnich:2009se}
\bibitem{Barnich:2009se}
  G.~Barnich and C.~Troessaert,
  ``Symmetries of asymptotically flat 4 dimensional spacetimes at null infinity revisited,''
  Phys.\ Rev.\ Lett.\  {\bf 105} (2010) 111103
    [arXiv:0909.2617 [gr-qc]].
  %%CITATION = doi:10.1103/PhysRevLett.105.111103;%%

%\cite{Barnich:2010eb}
\bibitem{Barnich:2010eb}
G.~Barnich and C.~Troessaert,
``Aspects of the BMS/CFT correspondence,''
JHEP \textbf{05} (2010), 062
doi:10.1007/JHEP05(2010)062
[arXiv:1001.1541 [hep-th]].

%\cite{Campiglia:2014yka}
\bibitem{Campiglia:2014yka}
M.~Campiglia and A.~Laddha,
``Asymptotic symmetries and subleading soft graviton theorem,''
Phys. Rev. D \textbf{90} (2014) no.12, 124028
doi:10.1103/PhysRevD.90.124028
[arXiv:1408.2228 [hep-th]].

%\cite{Campiglia:2015yka}
\bibitem{Campiglia:2015yka}
M.~Campiglia and A.~Laddha,
``New symmetries for the Gravitational S-matrix,''
JHEP \textbf{04} (2015), 076
doi:10.1007/JHEP04(2015)076
[arXiv:1502.02318 [hep-th]].

%\cite{Peraza:2023ivy}
\bibitem{Peraza:2023ivy}
J.~Peraza,
``Renormalized electric and magnetic charges for $O(r^n)$ large gauge symmetries,''
[arXiv:2301.05671 [hep-th]].


\end{thebibliography}
\end{document}